\documentclass[twocolumn,showpacs,superscriptaddress,preprintnumbers,amsmath,amssymb,longbibliography,pre]{revtex4-1}
\usepackage{dcolumn}
\usepackage{bm}
\usepackage{amsmath}
\usepackage{color}
\usepackage{wrapfig}
\usepackage{lipsum}

\usepackage{graphicx} 
\graphicspath{{./figures/}} 
\DeclareGraphicsExtensions{.pdf} 

\usepackage{url} 

\usepackage[utf8]{inputenc} 
\usepackage{amsmath} 
\usepackage{amsfonts} 

\usepackage{algorithm} 
\usepackage{algpseudocode} 

\usepackage{tabularx} 
\usepackage{booktabs} 
\usepackage{multirow} 

\usepackage{amsmath,amsthm,amsfonts,txfonts}
\usepackage{color}

\newcommand{\eqdef}{\coloneqq}     
\DeclareMathOperator{\Prob}{P}            

\begin{document}

\title{Scaling and Trade-offs in Multi-agent Autonomous Systems}
\author{Abram H. Clark}
\affiliation{Department of Physics, Naval Postgraduate School, Monterey, CA, 93943.}%
\author{Liraz Mudrik}
\affiliation{Department of Mechanical and Aerospace Engineering, Naval Postgraduate School, Monterey, CA, 93943.}
\author{Colton Kawamura}
\affiliation{Department of Physics, Naval Postgraduate School, Monterey, CA, 93943.}%
\author{Nathan C. Redder}
\affiliation{Department of Physics, Naval Postgraduate School, Monterey, CA, 93943.}
\author{Jo\~{a}o P. Hespanha}
\affiliation{Department of Electrical and Computer Engineering, University of California at Santa Barbara, Santa Barbara, CA 93106.}
\author{Isaac Kaminer}
\affiliation{Department of Mechanical and Aerospace Engineering, Naval Postgraduate School, Monterey, CA, 93943.}

\begin{abstract}
Designing autonomous drone swarms is hampered by a vast design space spanning platform, algorithmic, and numerical‑strength choices. We perform large‑scale agent‑based simulations in three canonical scenarios: swarm‑on‑swarm battle, cooperative area search with attrition, and pursuit of scattering targets. We demonstrate how dimensional‑analysis and data‑scaling can be leveraged to collapse performance data onto scaling functions that are mathematically simple, yet counterintuitive and therefore difficult to predict a priori. These scaling laws reveal success-failure boundaries, including sharp break points which we show can be framed as an ``effective swarm size.'' Additionally, we show how this technique can be used to quantify trade‑offs between agent count and platform parameters such as velocity, sensing or weapon range, and attrition rate. Furthermore, we show the benefits of embedding an optimal path planning loop within this framework, which can qualitatively improve the scaling laws that govern the outcome. The methods we demonstrate are highly flexible and would enable rapid, budget‑aware sizing and algorithm selection for large autonomous swarms.

\end{abstract}

\maketitle


\section{Introduction}

Drone swarms are groups of autonomous vehicles that coordinate to achieve goals~\cite{kallenborn_2020}. Technological improvements in computer processing, drone-to-drone communication, and energy storage density~\cite{kallenborn_2018} have made larger swarms more plausible. However, designing and fielding large swarms requires answers to many design problems, such as which drones would be most suitable, the best cooperative algorithm to use, and the minimum number of drones required. Financial considerations may also play some role. 


Thus, drone-swarm design naturally involves questions about scaling and trade-off behavior to determine optimal algorithms or platform specifications---such as speed, weapon range, or communication range---or to address financial or technological limitations associated with different platform specifications. For example, a fixed budget may allow the construction of many slow, cheap drones or a few fast, expensive drones. Algorithm $A$ may be found to perform best for the large swarm of slow drones, and algorithm $B$ may work best for the small swarm of fast drones, with an overall better performance for the small swarm of fast drones using algorithm $B$. Therefore, designing and fielding an optimal drone swarm requires mapping out how some performance metric depends on a potentially large number of parameters as well as discrete algorithmic choices.

Additionally, the dependence of the performance metric on any individual parameter might be complicated and not easily guessed~\cite{hamann2020guerrilla,ligot2022using,debie2023swarm}. For instance, the performance may include a ``scaling break point,'' where the change in the number of drones in a swarm leads to a saturation or even a decrease in the performance~\cite{hamann2020guerrilla}. For example, crowding may inhibit performance in a foraging problem~\cite{lerman2002mathematical}. Another example is minimum resource allocation, e.g., the minimum number of drones required to succeed in an adversarial engagement~\cite{navy_drone}. As we show, predicting where these scaling break points occur is often impossible using intuition or simple calculation. These scaling break points will likely vary with other parameter values, such as drone speed or sensing range.


In this paper, we demonstrate how dimensional analysis and data-scaling---techniques that are well established in physics, fluid mechanics, and other complex systems sciences~\cite{buckingham1914,barenblatt1987dimensional,gibbings2011dimensional,buckpi}---can be leveraged to obtain functions that relate the performance of a swarm to all the underlying parameter values. These functions are often mathematically simple, but they involve emergent functional forms (like power-law functions with nontrivial exponents) and are therefore difficult or impossible to guess. We also show how this approach naturally answers the questions of scaling and trade-off behavior above, including the location of scaling break points and how they depend on other parameter values. { Such scaling from local interaction rules to macroscopic behavior is common in active matter and collective motion~\cite{vicsek1995novel,marchetti2013hydrodynamics,ramaswamy2010mechanics,bechinger2016active}. While classic models in these fields primarily characterize spontaneous ordering, here we apply a similar approach to adversarial swarm engagements.}

Although the technique we show can be used on real-world data, we use computer simulations to generate large data sets involving different algorithms and wide variation in input parameters. We study three different scenarios: one, where red and blue swarms both try to kill each other using simple swarm rules; two, where a swarm of agents searches an area under the presence of attrition (via random failures or some active threat); and three, where a swarm must visit some number of scattering targets. We emphasize that the scaling laws derived from these simulations are mathematical properties of the models themselves; validating the underlying dynamics against hardware experiments is an important but separate research question.

Finally note that some of the present authors' previous works have approached swarm engagements from the perspective of an optimization problem~\cite{tsatsanifos2021modeling,walton2022defense,park2018observability,walton2018optimal,tsatsanifos2020computationally}. These studies assumed specific models for dynamics and weapons and showed that optimal trajectories could be formulated using direct methods of optimal control. This is done by formulating a performance metric (e.g., the survival probability of one's own drones or the total time needed to accomplish one's goals) and then using standard methods to find drone trajectories that maximize the performance metric. In these studies, the drones were not responding in real-time to an evolving scenario but flying pre-planned trajectories as determined by the optimization algorithm. We apply optimization to one of our three case studies, showing that (1) scaling analysis can still be applied to analyze the performance of optimal paths and (2) optimization can categorically improve the performance of a swarm.

\section{Data Scaling for Swarm Performance}
\label{sec:scal-anal}

In this section, we describe the basic problem of obtaining a performance metric and discuss how dimensional analysis and data-scaling can be applied. We start by assuming that, for a given algorithm and certain type of drone, there is a quantifiable performance metric $P$ that depends on all continuously varying parameters, which represent platform specifications of the drones. For example, for algorithm $A$ and parameters $p_1$, $p_2$, etc., then we can in principle write $P = f_A(p_1,p_2,...)$, where $f_A$ is some function to be determined. For algorithm $B$, there may be some different function such that $P = f_B(p_1,p_2,...)$. These functions may or may not bear any resemblance to each other.

Determining the function that relates $P$ to the inputs $\{p_n\}$ appears daunting. Systems comprised of large numbers of interacting objects (like drone swarms) often display \textit{emergent} behavior, meaning behavior that is not obvious or easily guessed from the individual components or their interactions~\cite{pines2016emergent,zottl2016emergent,kelley2013emergent}. Thus, mathematically relating performance metrics to input parameters requires running a very large number of scenarios, where all parameters are varied and the performance metrics are measured. If the drone behavior is deterministic, then $P$ can be measured for many combinations of $\{p_n\}$. If the drone behavior includes some randomness, then $P$ can still be obtained by ensemble averaging, assuming the underlying probability distributions can be approximated. This is often infeasible experimentally, but computer simulations are well suited for such a task. The results can be compared with experimental ground-truth. In this study, we use computer simulations, but our approach here can be easily applied to experimental data assuming it can be obtained.

Therefore, assuming $P$ can be computed for many scenarios, varying the parameter values, the remaining task is to obtain or at least approximate the function $P = f(\{p_n\})$. Such a task arises often in physical sciences, such as in fluid mechanics or soft matter physics. The form of the function $f$, as well as the way the parameters $\{p_n\}$ appear in the function, often gives significant insight into the emergent features of the problem. The point of this paper is to demonstrate how this function $f$ can be found for drone simulations, guided by dimensional analysis, in order to answer many of the issues raised in the introduction (e.g., trade-offs and scaling break-points).

The Buckingham-$\pi$ theorem makes a mathematical guarantee about the maximum number of unique arguments to $f$~\cite{buckingham1914,buckpi,evans1972dimensional,gibbings2011dimensional,barenblatt1987dimensional}. $P = f(\{p_n\})$ must depend only on dimensionless combinations of parameters, called dimensionless groups or $\pi$ groups. The maximum number of independent dimensionless groups is {  $k=n-D$,} where $n$ is the total number of parameters and {  $D$} is the number of physical dimensions (e.g., {  $D=3$} for mass, length, and time). There should then be some functional relation among the $\pi$ groups:
\begin{equation}
    \pi_1 = f(\pi_2,...,\pi_k).
\end{equation}
This method is common in fluid mechanics~\cite{rott1990note,garaud2021journey}, granular flow~\cite{dacruz2005,Krizou2020}, and other nonlinear dynamics problems~\cite{strogatz2018nonlinear}. 


The Buckingham-$\pi$ theorem thus provides a guide in searching for mathematical relations between input and output parameters for a complex process. The function $f(\pi_2,...,\pi_k)$ often turns out to be easy to interpret, and, while not guaranteed by the theorem, many of the inputs $\pi_2$ through $\pi_k$ can often be combined and further simplified or even neglected. This often requires trial-and-error combined with some intuition, as we show in the case studies below.

Adversarial swarm engagements necessarily involve a number of parameters involving physical units like mass $M$, length $L$, and time $T$ (e.g., velocity has units of $LT^{-1}$) or without units (e.g., number of drones or probability parameters have no units, $-$). In the examples we study here, we will use: numbers of drones $N$ ($[N] = -$); velocities $V$ ($[V] = LT^{-1}$); ranges $R$ for sensors or weapons ($[R] = L$); rates $\lambda$ for either weapons or some other kind of attrition ($[\lambda] = T^{-1}$); and an acceleration time scale $\tau$ ($[\tau] = T$). Here, $[A]$ means ``the units of $A$", and subscripts may refer to different kinds of drones (e.g., $N_d$ will be the number of defenders). The scenario in question should also have some kind of performance metric, such as a total time $t_{\rm tot}$ ($[t_{\rm tot}] = T$) or the average success probability $P$ ($[P] = -$). Dimensionless parameters like $N$ or $P$ form their own $pi$-groups, and dimensioned parameters like $V$ or $\tau$ should be combined into dimensionless ratios.

One interesting feature that is not guaranteed by the theorem, but that appears in the scenarios we consider is that a performance metric can be defined for $\pi_1$, and then $\pi_1$ can be written as function of a single quantity that depends on the size of the swarm $N$ as well as all other $\pi$ groups. As we show, this quantity can be interpreted as an \textit{effective size of the swarm}. When this quantity is large, the swarm is successful; when the quantity is small, the swarm is unsuccessful. Additionally, the functional form of $f$ as well as the way that all $\pi$ groups appear in $f$ is emergent and not easily guessed from the basic rules of the simulations. Instead, it arises from complex, many-body interactions during the engagements, which similarly occurs in groups of animals~\cite{kelley2013emergent,koorehdavoudi2016statistical}. 

\section{Methods}

We consider three case studies, in which we define a performance metric and vary multiple parameters to obtain $P(\{p_n\})$. We then consider how to form dimensionless $\pi$-groups from the set of $p_n$. We show how the data can be collapsed onto a single curve, relating $P$ to some fairly simple function of all the parameters. 

The first case study involves two adversarial swarms attacking each other. A red (attacking) swarm of $N_a$ agents flies directly at a high-value unit. A blue (defending) swarm of $N_d$ agents is launched from the high-value unit to intercept the red swarm. Each agent is equipped with weapons, which we model as simple attrition-rate functions that depend on the relative distance between any given pair of red and blue agents. Agents are killed randomly using Monte Carlo methods. We define the performance metric $P$ as the ensemble-averaged probability of survival of any individual agent from the red swarm ($P=0$ means blue wins, $P>0$ means red wins). We vary the rates of fire, $\lambda_a$ and $\lambda_d$, and the ranges, $R_a$ and $R_d$ of the weapons, and we find that a straightforward (but non-intuitive) formula can be found for $P$ in terms of $R_a/R_d$ and $\lambda_a/\lambda_d$, which can then be used to determine the minimum size of the blue swarm required for a successful mission.

The second case study is a search problem, where a group of $N$ agents coordinate to search a given area $A$. The agents divide up the search area into different sub-sectors and search their sub-sector before returning to base. We include random attrition of agents with a constant attrition rate $\lambda$, which could be from an active threat or simply due to random failures. We study cases with and without communication, where agents can intermittently share data with nearby agents. We define the performance metric $P$ as the percent of the desired area that has been searched and had its data reported back to base. We vary the ranges of sensing, $R_s$, and communication, $R_c$, as well as $A$ and $\lambda$. We find that the presence of communication between agents qualitatively changes the scaling.

The third case study involves coordinated pursuit of $N_a$ ``attacking'' targets by $N_d$ ``defending'' agents. All targets scatter in random directions with randomized velocities, maintaining a constant velocity and bearing in time. At least one pursuing agent must visit each target, and we define the performance metric $P$ as the total time required to visit all targets. We use an auctioning algorithm, where the target that is farthest from the point at which the targets originated is the highest priority and is assigned the closest defender; all other attackers are then assigned based on priority and nearest defender. We vary the numbers $N_a$ and $N_d$, the minimum range {  $R_d$} that a defender must approach an attacker to visit it, the velocity ratio {  $\nu$} of defenders relative to attackers, and the time scale $\tau$ for defenders to approach their asymptotic velocity. In this case of two swarms with competing but highly asymmetric goals, we also observe interesting nonlinear scaling in the parameters.

We use agent-based {  simulations} (meaning we resolve the position and velocity of every agent at all times), written in MATLAB, consisting of two components. First, the dynamics of each agent are modeled. In general, we do this by numerically integrating Newton's equation of motion. In the second case study (underwater search mission), all drone paths are preplanned, meaning there is no need to explicitly integrate a second-order differential equation. {  In} the general case, the forces on each agent consist of a thrust force toward a so-called ``virtual leader'', which is essentially the coordinates where the agent wants to go, as well as additional forces that arise {  from} interactions with other drones. The velocity of each agent is limited by a drag-like term, such that a maximum velocity $V_a$ or $V_d$ is asymptotically approached. Second, in addition to the dynamics of agents, we keep track of the set of surviving attackers $S_a$ and the set of surviving defenders $S_d$. At the beginning of all simulations, all $N_a$ attackers are included in $S_a$ and all $N_d$ defenders are included in $S_d$. Attackers and defenders are removed randomly from $S_a$ and $S_d$ based on attrition functions, as discussed below. 

The most general form for the equations of motion are, for attacker $i$
\begin{equation}
    m_i \ddot{\mathbf{r}}_i = K_i { \frac{\mathbf{r}_{vl}^{(i)}-\mathbf{r}_i}{|\mathbf{r}_{vl}^{(i)}-\mathbf{r}_i|}}  - B_i \dot{\mathbf{r}}_i  + \sum_{k \in S_a}\mathbf{F}_{aa}(\mathbf{r}_i-\mathbf{r}_k)
    + \sum_{l \in S_d}\mathbf{F}_{ad}(\mathbf{r}_i-\mathbf{r}_l),
    \label{eqn:eq-of-mot-att}
\end{equation}
or defender $j$
\begin{equation}
    m_j \ddot{\mathbf{r}}_j = K_j { \frac{\mathbf{r}_{vl}^{(j)}-\mathbf{r}_j}{|\mathbf{r}_{vl}^{(j)}-\mathbf{r}_j|}}  - B_j \dot{\mathbf{r}}_j + \sum_{k \in S_a}\mathbf{F}_{da}(\mathbf{r}_j-\mathbf{r}_k)
    + \sum_{l \in S_d}\mathbf{F}_{dd}(\mathbf{r}_j-\mathbf{r}_l)
    \label{eqn:eq-of-mot-def}
\end{equation}
{ Here, subscripts or superscripts $i$ and $j$ refer to the nominal attacker and defender; $k$ and $l$ refer to summed-over attackers and defenders (respectively); and dots denote time derivatives.} Additionally, $m$ is the inertial mass, $\mathbf{r}$ is position, $K$ is a thrust force magnitude, $\mathbf{r}_{vl}$ is the position of a virtual leader, and $B$ is damping parameter. $\mathbf{F}_{aa}(\mathbf{r}_i-\mathbf{r}_k)$ is a pseudo-force due to an interaction with the $k$-th attacker, and the sum of these pseudo-forces is interpreted by the drone as a thrust instruction. The other forces $\mathbf{F}_{ad}$, $\mathbf{F}_{da}$, and $\mathbf{F}_{dd}$ are similar. In each sum over $k$ and $l$, it is implied that the self-interaction force is set to zero for $i=k$ or $j=l$. We note that, neglecting these interaction forces, $V = K/B$ sets the typical velocity scale and $\tau = m/B$ sets a time scale for acceleration up to this velocity. Interaction pseudo-forces can cause the velocity to exceed the $V$. These equations represent a simplified dynamics model, but are sufficient to generate adversarial swarming engagements as well as demonstrate the scaling and trade-off techniques that are the focus of this paper. 

Equations~\eqref{eqn:eq-of-mot-att} and~\eqref{eqn:eq-of-mot-def} represent a system of $N_a+N_d$ second-order ordinary differential equations which may or may not be coupled together. At each time step, based on the values of the interaction forces and the position of the virtual leaders, the acceleration is calculated based on this equation, and we then numerically integrate using a modified velocity-Verlet algorithm with a time step $\Delta t$, which we choose as sufficiently small that it does not affect our results.

During the course of each time step in the simulation, we compute the survival probability $Q_i$ for attackers and $P_j$ for defenders over that time step according to attrition rate functions $\phi$. So, during a time step $\Delta t$, {  the probability of survival $P^{\rm surv}_m$ of an arbitrary agent $m$ is
\begin{equation}
    { P^{\rm surv}_m = \prod_{n} [1-\phi^{(mn)}(\mathbf{r}_m-\mathbf{r}_n;\lambda_n,R_n)\Delta t]} \label{eqn:prob-surv-agent},
\end{equation}
where the product runs over all sources of attrition $n$, $\phi^{(mn)}$ is the attrition rate on agent $m$ due to source $n$, and $\lambda_n$ and $R_n$ are the rate of fire and range of that source. Equations~\eqref{eqn:prob-surv-att} and~\eqref{eqn:prob-surv-def} below are the two specializations of Eq.~\eqref{eqn:prob-surv-agent} used in the first and third case studies.} For the second case study (underwater search), there is only one source of attrition.

For the first and third case studies, we use pairwise attrition functions $\phi_{ad}(\mathbf{r}_i-\mathbf{r}_l)$, which specifies the attrition rate for attacker $i$ due to defender $l$, or $\phi_{da}(\mathbf{r}_j-\mathbf{r}_k)$, which specifies the attrition rate for defender $j$ due to attacker $k$. Both of these attrition functions depend strictly on the instantaneous relative position of the two drones in question. So, the probability of survival $Q_i$ of attacker $i$ is 
\begin{equation}
    Q_i = \prod_{l\in S_d} [1-\phi_{ad}^{(il)}(\mathbf{r}_i-\mathbf{r}_l;\lambda_d,R_d)\Delta t] \label{eqn:prob-surv-att},
\end{equation}
and the probability of survival $P_j$ of defender $j$ is 
\begin{equation}
    P_j = \prod_{k \in S_a} [1-\phi_{da}^{(jk)}(\mathbf{r}_j-\mathbf{r}_k;\lambda_a,R_a)\Delta t] \label{eqn:prob-surv-def}.
\end{equation}
These functions $\phi$ will depend on the relative distance between enemies as well as the range $R$ and rate of fire $\lambda$. For every agent at each time step, we generate a pseudo-random number in MATLAB using a uniform distribution between 0 and 1. If this number is greater than the survival probability $Q_i$ or $P_j$, then that attacker or defender is removed from the simulation (i.e., from $S_a$ or $S_d$), including all interactions in Eqs.~\eqref{eqn:eq-of-mot-att} through \eqref{eqn:prob-surv-def}. The forms for {  $\phi_{ad}$ and $\phi_{da}$} are specified below, depending on the scenario.

\section{Case Study 1: Swarm Collision with Mutual Attrition of Red and Blue}

\subsection{Description of scenario and equations}

The first scenario we consider involves a collision of a red swarm and blue swarm, both governed by Leonard dynamics~\cite{leonard2001virtual,ogren2004cooperative,walton2022defense} involving cohesive interswarm potentials, further explained below. The engagement begins with the red swarm some distance $L$ away from the high-value unit (HVU), where each red agent has a virtual leader centered at the HVU. The blue swarm, which is centered at the HVU, senses the red swarm and begins flying straight toward the center of mass of the red swarm. The red swarm agents also generate a pseudo-force that depends on the defenders' positions, causing them to fly around the defending blue agents. All agents are equipped with weapons, modeled by smooth attrition functions, with adjustable rates of fire $\lambda_a$ and $\lambda_d$ (i.e., overall magnitude) and ranges $R_a$ and $R_d$ (i.e., a scale factor related to the distance between the two enemy drones). The goal of this case study is to understand trade-offs between the number of drones $N_a$ and $N_d$ and the weapons parameters $R_a$, $R_d$, $\lambda_a$, and $\lambda_d$.

Mathematically, we solve Eq.~\eqref{eqn:eq-of-mot-att} for each attacker $i$, with $\mathbf{r}_{vl}^{(i)} = 0$ (the position of the HVU) and $K_i = B_i = m_i = 1$. The intraswarm force {  $\mathbf{F}_{aa}$} between attackers $i$ and {  $k$} is set to be~\cite{leonard2001virtual,ogren2004cooperative,walton2022defense}
\begin{equation}
\mathbf{F}_{aa} = \begin{cases}
        \frac{\mathbf{r}_k-\mathbf{r}_i}{|\mathbf{r}_i-\mathbf{r}_k|^2}\left(1-\frac{d_0}{|\mathbf{r}_i-\mathbf{r}_k|} \right) &\text{, if }|\mathbf{r}_i-\mathbf{r}_k| < d_1, \\
        0 &\text{, if }|\mathbf{r}_i-\mathbf{r}_k| \geq d_1.
        \end{cases}
        \label{eq:F_aa_att_case1}
\end{equation}
The first term in this equation is an attractive term that pulls attacker $i$ toward attacker {  $k$ if $d_0<|\mathbf{r}_i-\mathbf{r}_k| < d_1$;} we set $d_0 = 2$ and $d_1 = 3$. The second term is a repulsive term that pushes attacker $i$ away from attacker {  $k$ if $|\mathbf{r}_i-\mathbf{r}_k| < d_0$.} For the attackers, we also include a force from defender {  $l$}
\begin{equation}
\mathbf{F}_{ad} = \begin{cases}
        \frac{\mathbf{r}_l-\mathbf{r}_i}{|\mathbf{r}_i-\mathbf{r}_l|^2}\left(1-\frac{d_r}{|\mathbf{r}_i-\mathbf{r}_l|} \right) &\text{, if }|\mathbf{r}_i-\mathbf{r}_l| < d_r, \\
        0 &\text{, if }|\mathbf{r}_i-\mathbf{r}_l| \geq d_r.
        \end{cases}
        \label{eq:F_ad_att_case1}
\end{equation}
This force is repulsive for $|\mathbf{r}_i-\mathbf{r}_l| < d_r$ and zero otherwise. We set $d_r = 6$, which causes the attackers to want to fly around the defenders with a characteristic range of 6 (arbitrary units).

Defender dynamics are also solved using Eq.~\eqref{eqn:eq-of-mot-def} for each defender $j$, with $\mathbf{r}_{vl}^{(j)}$ set as the average position of all the attackers (i.e., the center of the attacking swarm) and $K_j = B_j = m_j = 1$. The intraswarm force $\mathbf{F}_{dd}$ between defenders $j$ and $l$ is set to the same function as $\mathbf{F}_{aa}$,
\begin{equation}
\mathbf{F}_{dd} = \begin{cases}
        { \frac{\mathbf{r}_l-\mathbf{r}_j}{|\mathbf{r}_j-\mathbf{r}_l|^2}\left(1-\frac{d_0}{|\mathbf{r}_j-\mathbf{r}_l|} \right)} &\text{, if }{ |\mathbf{r}_j-\mathbf{r}_l|} < d_1, \\
        0 &\text{, if }{ |\mathbf{r}_j-\mathbf{r}_l|} \geq d_1,
        \end{cases}
        \label{eq:F_dd_def_case1}
\end{equation}
where $d_0=2$ and $d_1 = 3$ as before. There is no collision avoidance for defenders to attackers, i.e., {  $\mathbf{F}_{da} = 0$.}

Attacker and defender weapons are both modeled using a cumulative normal distribution function, $\Phi(z)=\int_{-\infty}^z \frac{1}{\sqrt{2\pi}}e^{-\zeta^2/2}d\zeta$. We model attrition from defender $l$ to attacker $i$ as
\begin{equation}
    \phi_{ad}^{(il)}= \begin{cases} 2\lambda_d\Phi\left(-\frac{|\mathbf{r}_i - \mathbf{r}_l|}{ R_d}\right)&\text{, if }|\mathbf{r}_i-\mathbf{r}_l| = \displaystyle\min_{k \in S_a} ({ |\mathbf{r}_k-\mathbf{r}_l|}), \\
    0 &\text{, otherwise.} \end{cases}
\end{equation} 
Note that the conditional statement is modeling \textit{targeting}, where each defender chooses to target one attacker based solely on which attacker is closest. Similarly, we model attrition from attacker $k$ to defender $j$ as
\begin{equation}
    \phi_{da}^{(jk)}= \begin{cases} 2{ \lambda_a}\Phi\left(-\frac{|\mathbf{r}_j - \mathbf{r}_k|}{ { R_a}}\right)&\text{, if }|\mathbf{r}_j-\mathbf{r}_k| = \displaystyle\min_{l \in S_d} (|\mathbf{r}_l-\mathbf{r}_k|), \\
    0 &\text{, otherwise,} \end{cases}
\end{equation} 
where now each attacker chooses only to target a single defender, depending on which is closest. Agents are randomly removed from the simulation based on the forms of these attrition functions in combination with Eqs.~\eqref{eqn:prob-surv-att} and \eqref{eqn:prob-surv-def}.

Initial conditions for the simulations are generated by placing a red swarm at a distance $L$ from the HVU, placing a blue swarm at the HVU, and setting all survival probabilities to 1. The specific positions of the red and blue agents are generated randomly at points surrounding the starting positions for the swarms, and, for each result, we take the average of an ensemble of five initial conditions. We quantify the result of the simulation by the survival probability of the attackers, defined as the ratio of the number of alive attackers at the end of the simulation to $N_a$. We note that the survival probability of the HVU can also be tracked using an equation similar to Eq.~\eqref{eqn:prob-surv-agent}. We have checked that our results are similar in this case. We choose $P_a$ because it is both a good proxy for the HVU survival and a more sensitive measure of the success of the blue swarm (i.e., the HVU survival is zero if all blue agents are defeated, regardless of how many red agents survive).

\begin{figure*}[t]
\includegraphics[width = \textwidth]{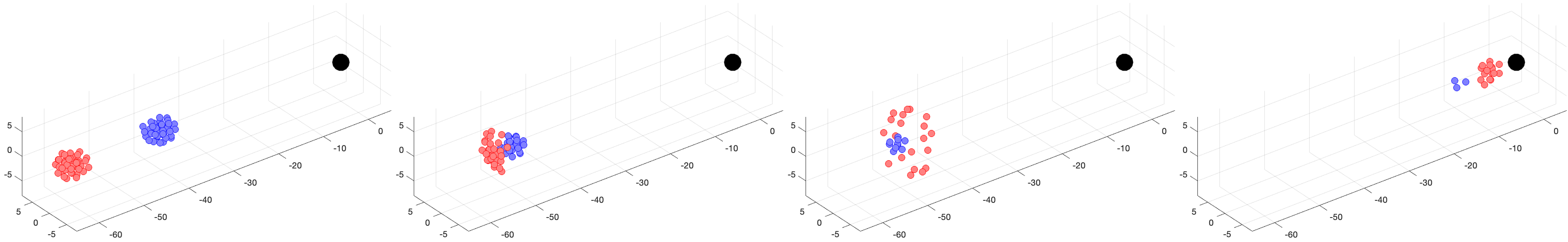}
\caption{\raggedright Four snapshots from an engagement where attackers destroy defenders as well as the HVU. Parameter values are $N_a = N_d = 40$, $\lambda_a = \lambda_d = 1$, and $R_a = R_d = 2 \sim d_r$.}
\label{fig:snapshots-case1} 
\end{figure*}

\subsection{Results and analysis}

Results from a sample engagement are shown in Fig.~\ref{fig:snapshots-case1}, with parameter values of $N_a = N_d = 40$, $\lambda_a = \lambda_d = 1$, and $R_a = R_d = 2$. The red swarm is flying toward the HVU (black circle) and is met by the blue swarm. The red swarm's collision avoidance causes it to spread out and fly around the blue swarm, and a tail-chase ensues. The engagement finishes with the red swarm destroying all blue agents as well as the HVU.

Despite the fact that red and blue agents are identical in weapons, dynamics, and intra-swarm interactions, the fact that there is a difference in the inter-swarm dynamics (red avoids blue, blue does not avoid red) provides an asymmetrical advantage for the red swarm. This advantage is not obvious to predict without observing the \textit{emergent} outcome of the engagement. Additionally, we note that the weapon ranges are similar to the $d_0$, $d_1$, and $d_r$, which makes clear intuitive predictions difficult. We now demonstrate how scaling analysis can be useful in quantifying trade-offs among $N_d$, $N_a$, $\lambda_d$, $\lambda_a$, $R_d$, and $R_a$ regarding the outcome of this engagement.

The first step is to identify a complete set of parameters that characterize an engagement. The output parameter to be studied is the probability of survival of an individual attacker, $P_a$, which can also be thought of as the ratio of the number of attackers surviving at the end of the engagement to the original number of attackers.
If $P_a>0$, then at least one attacker has made it to the base, and the mission has failed. We note that this parameter is also a $\pi$ group, since it is already dimensionless. The input parameters to be varied here are $N_d$, $N_a$, $\lambda_d$, $\lambda_a$, $R_d$, $R_a$; additionally, there are several parameters which will not be varied, which are $K_j$, $B_j$, $m_j$, $K_i$, $B_i$, $m_i$, $d_0$, $d_1$, and $d_r$. In total, these are 16 parameters, which can be used to write 13 dimensionless $\pi$ groups. The Buckingham-$\pi$ theorem only provides a guarantee that $P_a$ can be expressed as a function of the remaining 12 dimensionless numbers,
\begin{equation}
    P_a = f\left( N_d, N_a, { \frac{\lambda_a}{\lambda_d}}, \frac{R_d}{R_a}, ... \right). 
    \label{eqn:Buck-pi-example1}
\end{equation}

However, as discussed in Sec.~\ref{sec:scal-anal}, it is often true that some parameters can be further combined and simplified, and the functional form of $f$ can be expressed (at least in certain regions of the parameter space) using direct combinations or products of functions of the individual dimensionless groups. We find, in this case and the following case study, that the performance metric, in this case $P_a$, can be expressed as a function of $N_d/N_{a,{\rm eff}}$, where $N_{a,{\rm eff}}$ is an effective number of attackers which depends on both the number of attackers as well as the other parameters in the simulation. For example, increasing the attackers rate of fire causes $N_{a,{\rm eff}}$ to increase. 

The rest of this section proceeds as follows. First, we show that the functional form $P_a = f(N_d)$, with all other parameters fixed, is always well described by a curve that equals 1 for very small $N_d$ and then drops to 0 when $N_d$ crosses some critical value, which is different for each set of parameters. We then use curve fitting to extract the characteristic value of $N_d$ where $f$ goes to zero for all sets of parameters, and we define that value as $N_{a,{\rm eff}}$. We then consider how $N_{a,{\rm eff}}$ depends on all remaining parameters, finding that $N_{a,{\rm eff}}\propto N_a (\lambda_a/\lambda_d)^\alpha$ with $\alpha \approx 0.6$. The dependence on $R_a$ and $R_d$ is more complex and likely depends on $d_r$, which was not explicitly varied. This gives $P_a$ as a simple but not intuitive function of the numbers of drones, the rates of fire, and the ranges of the weapons.


\begin{figure}[h]
\raggedright (a) \\
    \includegraphics[width=0.99\columnwidth]{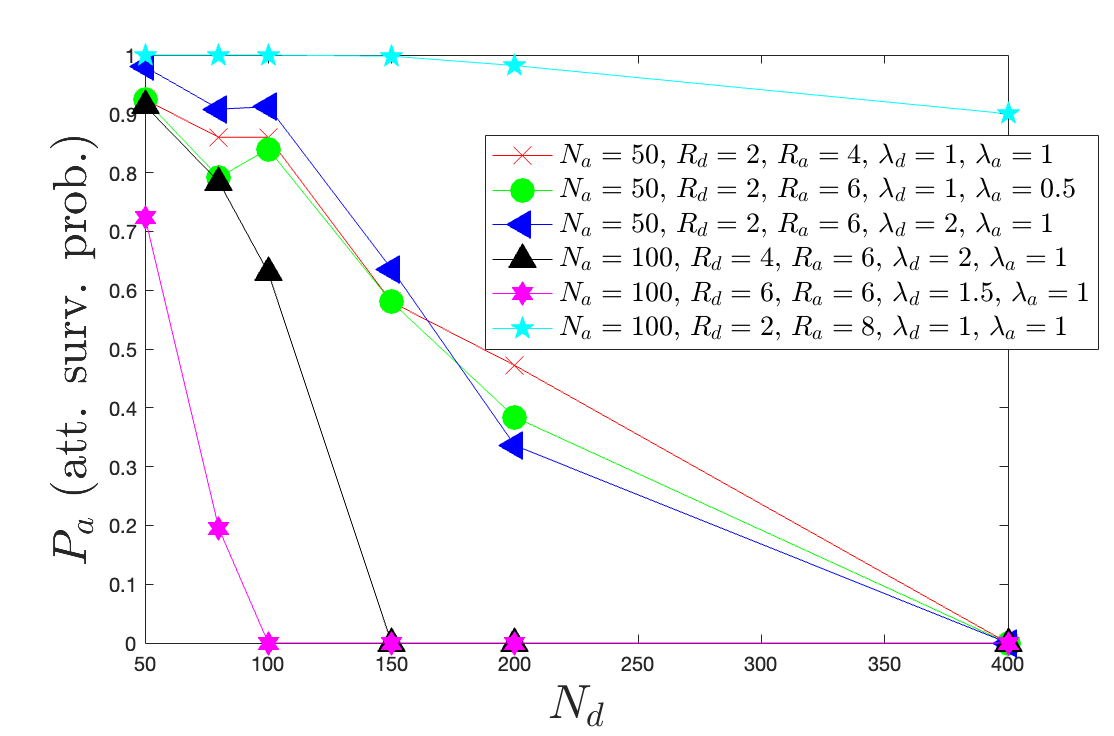} \\
    \raggedright(b) \\
    \includegraphics[width=0.99\columnwidth]{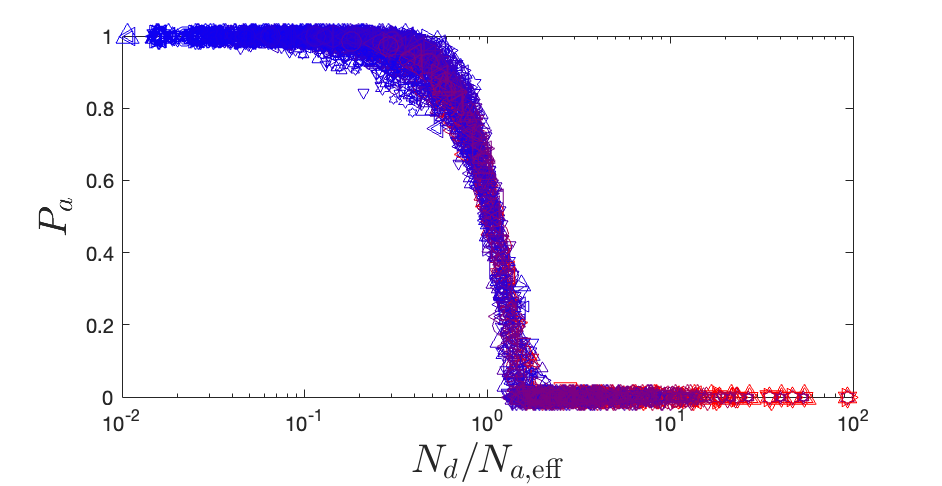}
    \caption{(a) Sample curves of $P_a$ versus $N_d$ for six combinations of other parameters. (b) $P_a$ versus $N_d/N_{a,{\rm eff}}$ for the curves shown in panel (a), plus many more different combinations of parameter values. The color scheme is strictly based on $R_a/R_d$. Note that $N_{a,{\rm eff}}$ is a different number for every curve.}
    \label{fig:Pa_vs_Nd}
\end{figure}

We first consider $P_a$ as a function of $N_d$, holding $N_a$, $\lambda_d$, $\lambda_a$, $R_d$, and $R_a$ fixed; sample curves are shown in Fig.~\ref{fig:Pa_vs_Nd}(a) for a few combinations of $N_a$, $\lambda_d$, $\lambda_a$, $R_d$, and $R_a$. Several features are immediately apparent from these curves. First, all curves look similar, in that they decrease from 1 to 0 as $N_d$ is increased, suggesting that all curves could be collapsed onto a master curve when plotted using the right combination of parameters. As we show, this provides a convenient and concise way to analyze the trade-offs inherent to this type of problem. Second, the blue leftward triangles and the green circles have identical parameters except $\lambda_a$ and $\lambda_d$, but with $\lambda_a/\lambda_d$ held constant. The resulting curves are nearly identical, confirming that this dimensionless ratio is dominant. Third, the red crosses ($\times$) are very similar, despite having two different kinds of parameters varied ($R_a$ and $\lambda_a$). This confirms that there are trade-offs between these parameters: in this case, increasing the range by a factor of 1.5 and decreasing the rate of fire by a factor of 2 yielded nearly identical results. 

The first feature discussed above, that all $P_a(N_d)$ curves can be collapsed onto a master curve, is demonstrated as follows. We perform simulations with all combinations of $N_d = 30$, 50, 80, 100, 150, 200, and 400; $N_a = 50$, 75, and 100; $\lambda_d = 0.5$, 1, 1.5, and 2; $\lambda_a = 0.5$ and 1; $R_d = 0.5$, 0.75, 1, 1.5, 2, 2.5, 3, 4, 6, and 8. We obtain similar curves to those shown in Fig.~\ref{fig:Pa_vs_Nd}(a), and we fit all data to a function of the form 
\begin{equation}
   P_a = \frac{1}{2}\left[1-\tanh \left( \log \frac{N_d}{N_{a,{\rm eff}}} \right)\right].
\end{equation}
The results of the analysis here are not sensitive to the form of this {  equation;} using the hyperbolic {  tangent} is not crucial, provided {  the function} drops from 1 to 0 as its argument increases. $N_{a,{\rm eff}}$ is unique to all other parameters, in general, but assuming it is known the data can be collapsed on the curve shown in Fig.~\ref{fig:Pa_vs_Nd}(b). If a functional form (or an approximate form) for $N_{a,{\rm eff}}$ were known in terms of all other parameters, then Eq.~\eqref{eqn:Buck-pi-example1} would be specified and the problem would be solved.

\begin{figure}[h]
    \includegraphics[width=0.99\columnwidth]{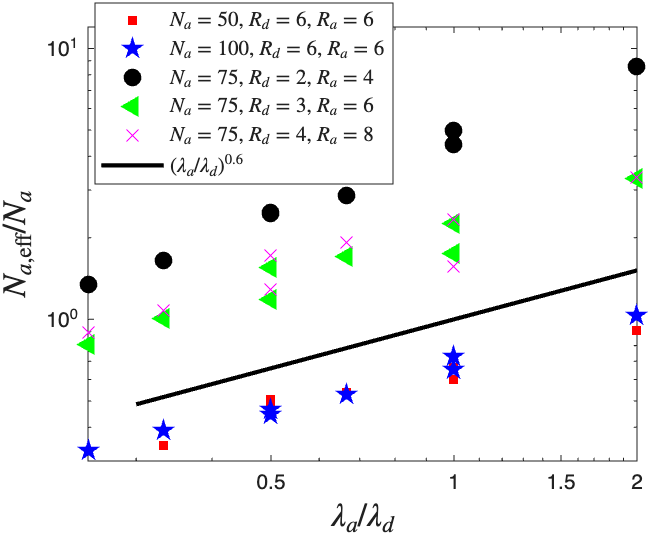} \\
    \caption{Plots of $N_{a,{\rm eff}}/N_a$ versus $\lambda_a/\lambda_d$ for five different combinations of parameter values. These curves demonstrate that $N_{a,{\rm eff}} \propto N_a (\lambda_a/\lambda_d)^\alpha$, with $\alpha \approx 0.6$.}
    \label{fig:Na_eff_scaling}
\end{figure}

To determine how $N_{a,{\rm eff}}$ depends on the remainder of the parameters, each one must be treated on a case-by-case basis. Guided by a Buckingham-$\pi$ approach, we make two assumptions, both of which are justified by the data in Fig.~\ref{fig:Na_eff_scaling} (as well as extensive other cases which are not shown). We first guess that $\lambda_a$ and $\lambda_d$ only affect $N_{a,{\rm eff}}$ as the dimensionless ratio $\lambda_a/\lambda_d$. We also guess that $N_{a,{\rm eff}} \propto N_a$ and only consider the dependence of the ratio $N_{a,{\rm eff}}/N_a$ on {  $\lambda_a/\lambda_d$.} Fig.~\ref{fig:Na_eff_scaling} shows that $N_{a,{\rm eff}}/N_a = A \left({ \frac{\lambda_a}{\lambda_d}}\right)^\alpha$, where $\alpha \approx 0.6$ and $A$ is an amplitude that depends on other parameters. A more precise value for $\alpha$, as well as confidence bounds, could be obtained with a more detailed analysis. The red squares and blue stars shown in Fig.~\ref{fig:Na_eff_scaling} show that $A$ is independent of $N_a$ (justifying the assumption that $N_{a,{\rm eff}} \propto N_a$). Additionally, multiple values of $\lambda_a$ and $\lambda_d$ are included for each specific symbol type (e.g., green triangles) in Fig.~\ref{fig:Na_eff_scaling}, but all data collapse when plotted as a function of the ratio $\lambda_a/\lambda_d$ (justifying the assumption that only this ratio affects $N_{a,{\rm eff}}$, not the two parameters individually). Thus, $A$ can be written as some function that depends on $R_a$ and $R_d$, i.e., $A(R_d,R_a)$.

Thus far, we have shown that $N_{a,{\rm eff}} \approx A(R_d,R_a) N_a { (\lambda_a/\lambda_d)}^{\alpha}$, where $\alpha \approx 0.6$. To characterize $A(R_d,R_a)$ we perform a best fit for $A$ in this approximate equation by taking the mean of all data for each combination of $R_a$ and $R_d$, denoted $A(R_d,R_a) \approx \left\langle \frac{N_{a,{\rm eff}}}{N_a} \left(\frac{\lambda_a}{\lambda_d}\right)^{-\alpha} \right\rangle$. This quantity is plotted in Fig.~\ref{fig:Na_eff_scaling-2}, where panel (a) shows $A$ as a function of the dimensionless ratio $R_d/R_a$ for the three different values of $R_a$ that we consider (4, 6, and 8). For $R_d/R_a > 0.6$, all curves collapse, demonstrating that $R_d/R_a$ captures the changes in $A$ for this regime. {  A best fit to the data in this regime returns $A(R_d/R_a) \approx 5 e^{-1.8 R_d/R_a}$. The black line shown in Fig.~\ref{fig:Na_eff_scaling-2}(a) shows exponential decay proportional to $e^{-1.8 R_d/R_a}$.

\begin{figure}[h]
    \raggedright(a) \\
    \includegraphics[width=0.99\columnwidth]{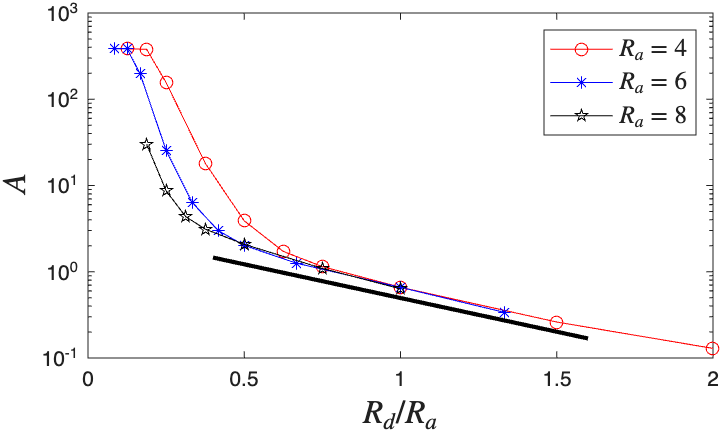} \\
    \raggedright(b) \\
    \includegraphics[width=0.99\columnwidth]{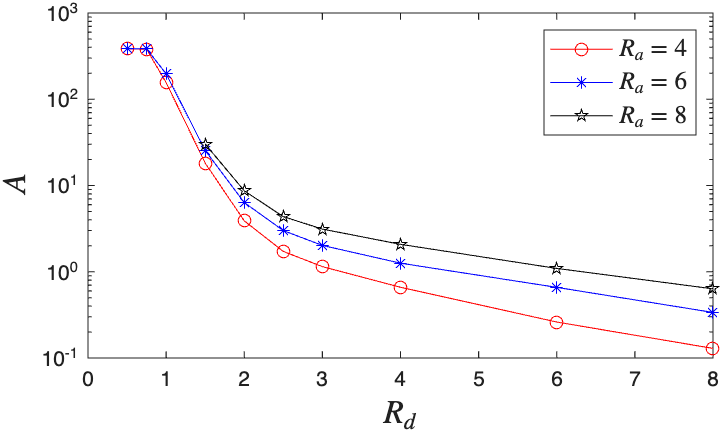}
    \caption{(a) $A$ versus $R_d/R_a$ for three values of $R_a$, where $N_{a,{\rm eff}} \propto A$ (see text). (b) $A$ versus $R_d$ for the same data.}
    \label{fig:Na_eff_scaling-2}
\end{figure}

When $R_d/R_a$ is near $1$, the curves in Fig.~\ref{fig:Na_eff_scaling-2}(a) collapse well. However, when $R_d/R_a \lesssim 0.6$, these curves separate from each other, meaning that $R_d/R_a$ is not the best dimensionless quantity in this regime. Fig.~\ref{fig:Na_eff_scaling-2}(b) shows the same data plotted simply as a function of $R_d$. This plot demonstrates that the rise in $N_{a,{\rm eff}}$ for small $R_d$ occurs independent of $R_a$. The value of $R_d$ where this rise occurs may vary if swarm-interaction parameters were varied (e.g., $d_r$).}

To summarize, Fig.~\ref{fig:Pa_vs_Nd} shows that the success metric $P_a$ (which we want to be zero, since all attackers should be killed) can be collapsed onto a single curve, where $P_a = 0$ for $N_d \gg N_{a,{\rm eff}}$ and $P_a = 1$ for $N_d \ll N_{a,{\rm eff}}$, with a crossover point at $N_d \approx N_{a,{\rm eff}}$. {  Guided by Buckingham-$\pi$, we find that $N_{a,{\rm eff}} \approx A\,N_a (\lambda_a/\lambda_d)^{\alpha}$, where $A \approx 5 e^{-1.8 R_d/R_a}$ when $R_d/R_a \gtrsim 0.6$. When $R_d$ is made small, $A$ instead rises by two to three orders of magnitude, independent of $R_a$.}

This form can be used to compare variation in each dimensionless parameter. For example, increasing the number of drones has a stronger effect (linear) than increasing the rate of fire (sublinear power law). Additionally, varying the ranges has a similar effect (albeit with a different functional form) to changes in the number of drones or the rate of fire. However, for sufficiently small {  $R_d$,} we find $N_{a,{\rm eff}}$ rises sharply, multiplying the {  effective} size of the {  attacking} swarm by a factor of 10 to 100. This means that there is a critical weapons range that is required for defenders, likely related to the avoidance distance {  $d_r$} of the attackers, which must be met before the defending swarm can be effective. {  Since we hold $d_r$ fixed here, we state this criterion in terms of $R_d$ rather than as a dimensionless threshold; establishing the latter would require a separate sweep over the swarm-interaction length scales.} The functions that model the attrition due to the weapons are all smooth, so this critical range requirement is not obvious \textit{a priori}. Finally, we emphasize that this analysis is limited to the specific scenario laid out here (i.e., the differential equations governing the flight dynamics and the attrition modeling); however, the approach is straightforward to apply to any scenario where two swarms are engaged in an adversarial contest.

To illustrate the practical utility of the scaling law derived above, consider a mission planner facing an attacking swarm of $N_a = 50$ red agents with weapons parameters $\lambda_a = 1$ and $R_a = 6$. Two defending platform options are available: Option~A offers $\lambda_d = 2$ and $R_d = 4$ at a unit cost of $C_A$; Option~B offers $\lambda_d = 1$ and $R_d = 6$ at a unit cost of $C_B$. For each option, the planner computes $N_{a,{\rm eff}} = N_a { (\lambda_a/\lambda_d)}^\alpha A(R_d/R_a)$ using the fitted formula and the values of $\alpha$ and $A$ from this case study. The minimum number of defenders required for mission success is $N_d \approx 2\,N_{a,{\rm eff}}$, so the total procurement cost for each option is $2\,N_{a,{\rm eff}} \times C_{A,B}$. Evaluating these expressions is instantaneous once the scaling law has been derived, replacing the need for additional large-scale simulation sweeps for each design alternative.

\section{Case Study 2: Underwater Search Mission}

\subsection{Scenario Description}
We consider an underwater search mission using $N$ cooperating autonomous underwater vehicles (AUVs) equipped with sonar with sensing range $R_{s}$. The AUVs have maximum battery time of $B$ and maximum speed of $V$. We assume attrition, either by direct threats, random mechanical failures, or some other mechanism. Thus, some of the AUVs will be lost, and the area searched by any lost AUV is not considered searched, unless it is shared with another AUV and thus reported back to base. Our central question is whether and to what degree communication between drones improves the total coverage of the area to be searched. We find that the presence of communication provides a significant improvement, and we demonstrate how the two curves (with and without communication) scale with $N$, $R_s$, $V$, attrition rate $\lambda$, total search area $A$, and communication range $R_c$ (assumed to be small to reduce the probability of detecting the AUVs).

\begin{figure}
    \centering
    \includegraphics[width=\linewidth]{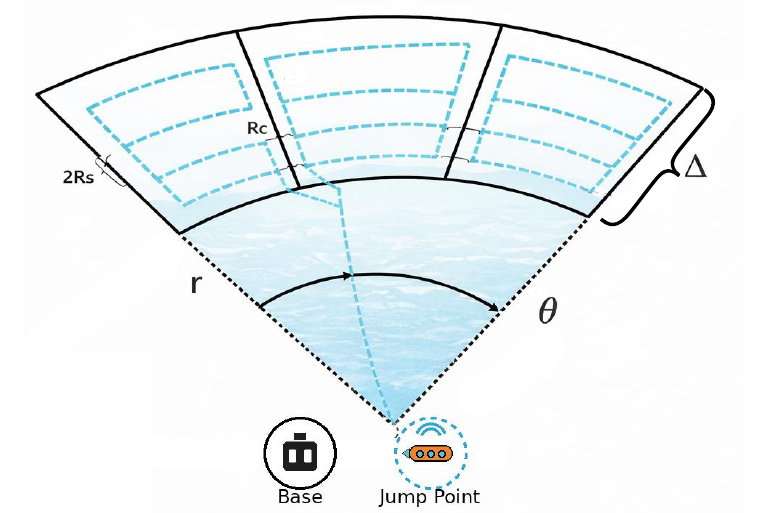}
    \caption{Schematic overview of a cooperative underwater search mission.}
    \label{fig:schematic}
\end{figure}

Fig.~\ref{fig:schematic} presents a schematic overview of executing a cooperative underwater search mission using three AUVs. At the beginning of the scenario, the AUVs are placed at their base, some distance $r$ from the search area {  $A$, determined} by angle $\theta$ and depth $\Delta$. The AUVs then travel together from the base into a jump point. Each AUV is assigned an equal portion of the search area, and at this point, they approach their subsection. Each AUV scans its subsection using a ``lawnmower'' pattern, as shown in Fig.~\ref{fig:schematic}. We assume all communication is done at the boundaries between subsections, when neighboring drones are closest together. The point of closest approach is set by $R_c$, and the width between sweeps is equal to $2R_s$, which guarantees the entire area is covered. Once the AUVs complete the search mission, they return to the jump point separately and then together to the base.

\begin{figure}[h] 
\centering
\includegraphics[width=0.99\columnwidth]{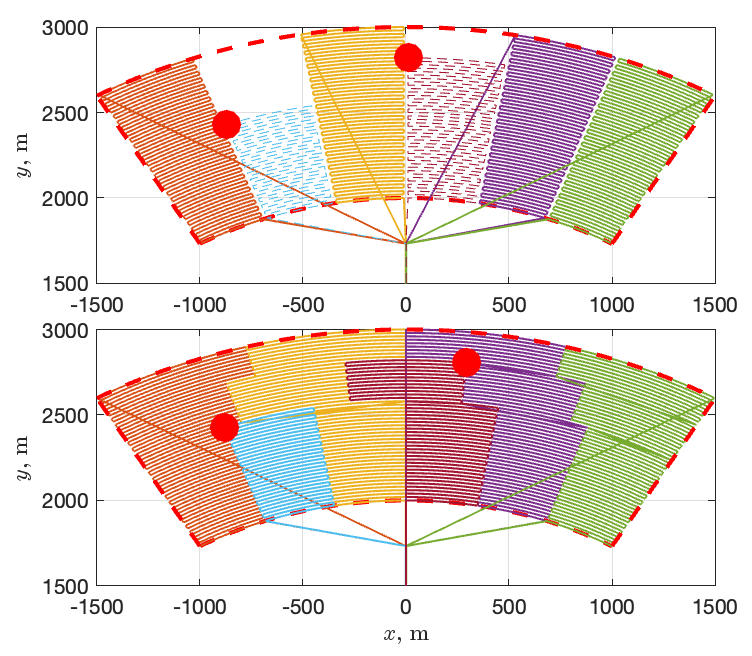}
\caption{Illustration of attrition effect over the search pattern. (a) Without information sharing. (b) With information sharing.}
\label{fig:Death} 
\end{figure}

If battery life is unlimited and there is no attrition of any kind, then the search area will always be completely explored. However, in real-life scenarios, the AUVs are exposed to different threats.
For example, in mine countermeasures missions, the AUVs might trigger the mines, or the wildlife might attack them when exploring the bottom of the ocean.
Therefore, we also consider the possibility of an unexpected event that causes the death of one or more AUVs during the search mission.
We model the event of losing an AUV as a Poisson process with parameter $\lambda$.
If the AUVs do not share information, they cannot adapt to the unexpected changes in their nominal mission, as presented in Fig.~\ref{fig:Death}(a), where the perfect coverage performance is reduced to $66\%$ in the presence of threats.
However, when information is shared between the AUVs, they are able to adapt to a loss and rearrange their search pattern, as presented in  Fig.~\ref{fig:Death}(b). We assume that information sharing is instantaneous; in reality, there would be lags, where the fact that agent $j$ is lost would be communicated gradually to the entire fleet, moving out one subsection in each direction on successive sweeps.

\subsection{Results and analysis}

We first identify the complete set of parameters that characterize the search mission. We choose the mean percentage of covered area, $P_{a}$, as the performance metric. The input parameters to be varied here are $N$, $R_{s}$, $R_{c}$, $A$, and $\lambda$. $V$ is also used in the dimensional analysis but not directly varied. This makes seven total parameters ($P_a$, plus six input parameters). The parameters $r$, $\theta$, and $\Delta$ are the geometric properties of the search area and will not be varied directly. We choose $B$ sufficiently large that it does not affect our results (e.g., drones do not run out of batteries).

{  These} seven parameters can be used to write five dimensionless $\pi$ {  groups, since} only length and time dimensions are {  involved.} 
The Buckingham-$\pi$ theorem only provides a guarantee that {  $P_{a}$} can be expressed as a function of the remaining four dimensionless numbers: $N$, $R_c/R_s$, $A/{R_s}^2$, and $V/\lambda R_s$.
\begin{equation}
    { P_{a}} = f\left(N,\frac{R_c}{R_s}, \frac{A}{{R_s}^2}, \frac{V}{\lambda R_s} \right). 
    \label{eqn:Buck-pi-example3}
\end{equation}


In this case, each AUV has a battery life of $B = 10$~hr.
We vary the number of AUVs to be $4$, $6$, $8$, $10$, $12$, $14$, $16$, $18$, and $20$, the sensing range to be $5$, $10$, and $20$~m, the communication range for the information sharing case is set to be $1$ and $4$~m, and the longitudinal attrition rate, $\lambda / V$, to be $4\cdot10^{-5}$, $6\cdot10^{-5}$, $8\cdot10^{-5}$, and $10^{-4}$~1/m.
The search area is distanced $2$~km away from the base, $\Delta = 1$~km, and $\theta$ is set to be $30^{\circ}$, $60^{\circ}$, and $90^{\circ}$.
For each of these values, we perform 10,000 simulation runs to get the mean coverage percentage.
Therefore, we run $6.48$ million simulations for the information sharing case and $3.24$ million runs for the non-sharing case to get the following analysis.

\begin{figure}
    \centering
    \includegraphics[width=0.99\linewidth]{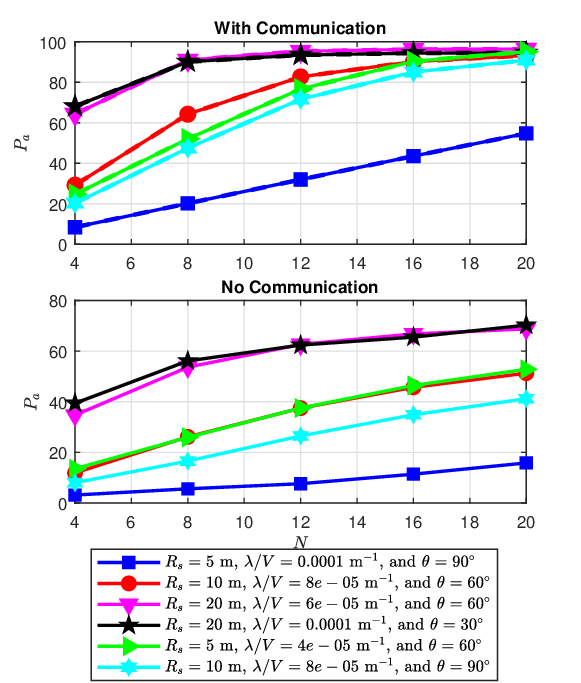}
    \caption{ Sample curves of $P_{a}$ versus $N$ for six combinations of the other parameters with (upper figure) and without (lower figure) communication capabilities.}
    \label{fig:Pa_vs_N}
\end{figure}

Fig.~\ref{fig:Pa_vs_N} shows selected curves for $P_a$ versus $N$ for selected values of $R_{s}$, $\lambda / V$, and $\theta$.
The upper graph shows the behavior when communication capabilities exist between the AUVs, and the lower graph shows the behavior when no such capability exists.
We also notice different behaviors for different curves. For example, the blue line seems to increase linearly in this range of $N$, while the other lines become saturated near $100\%$.
In the lower graph, we get significantly lower values for $P_{a}$ than in the upper graph due to the lack of communication capabilities.
The black and magenta curves are steeper than the rest, while the blue curve increases moderately. 
Still, in this case, none of the curves get saturated.

\begin{figure}
    \centering
    \includegraphics[width=0.99\columnwidth]{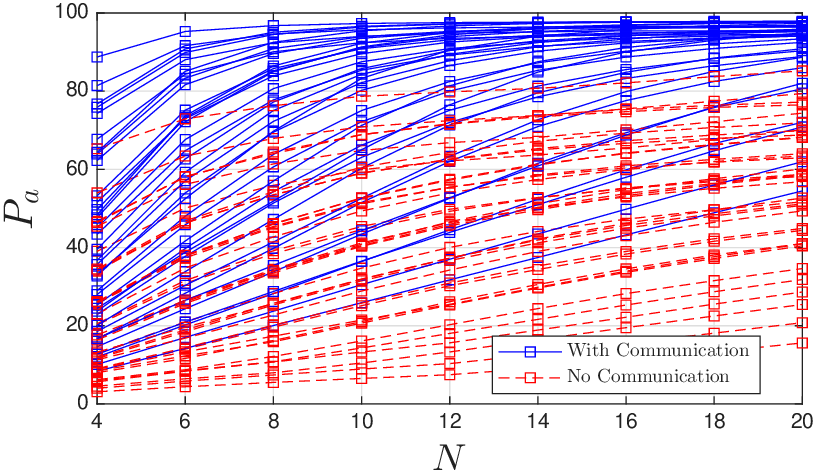} \\ 
    \includegraphics[width=0.99\columnwidth]{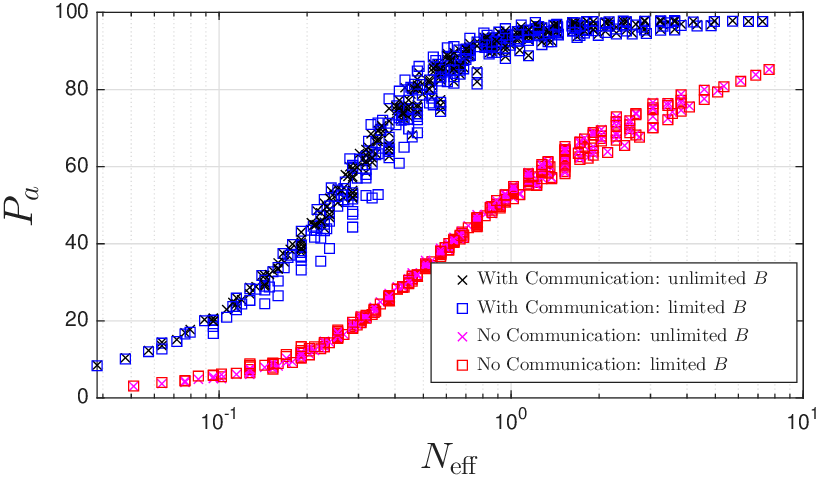}
\caption{The upper plot presents $P_a$ versus $N$ for all simulated cases with communication enabled (solid blue lines) and disabled (dashed red lines). The many curves shown correspond to variations in all parameters, the values of which are given in the main text, showing that the performance curves have a similar shape for all parameter combinations, but with significant horizontal shifts. The lower plot shows the same data, but with effective swarm size $N_{\mathrm{eff}}$ on the horizontal axis, demonstrating excellent collapse onto two separate curves: one with communication and one without. For both communication cases, we consider unlimited battery time (x markers) and limited battery time (square markers); battery limitations only affect performance in extreme cases involving small swarms covering large areas.}

    \label{fig:Pa_vs_Neff}
\end{figure}

The behavior of all simulations is shown in the upper plot of Fig.~\ref{fig:Pa_vs_Neff}, amounting to a total of $36$ curves for each case of using communication (solid blue lines) or not (dashed red lines).
This figure shows the different behaviors of the curves by using different scenario parameters.
We can also deduce the importance of adding communication capabilities, as we can clearly see the performance improvement that results from this.
However, no clear behavior can describe all the presented curves. 


On the other hand, the curves in the {  upper} plot of Fig.~\ref{fig:Pa_vs_Neff} do appear qualitatively similar in that they smoothly vary from 0 to {  $100\%$} as the number of drones is increased. This strongly suggests that they could be scaled onto a single curve if plotted in terms of the correct function of the parameters. {  The lower plot of} Fig.~\ref{fig:Pa_vs_Neff} shows such a scaled plot, where $P_a$ is plotted as a function of $N_{\rm eff}$, defined as
\begin{equation}
    N_{\text{eff}} = 
    \begin{cases}
        \frac{V R_{s}}{\lambda A} (N - 1), & \text{With com.} \\
        \frac{V R_{s}}{\lambda A} N, & \text{No com.}
    \end{cases}
\end{equation}
We note that this parameter is simply a product of the three dimensionless parameters from above, neglecting the communication range: $N$, $\frac{A}{{R_s}^2}$, and $\frac{V}{\lambda R_s}$. In particular, $N_{\rm eff} =\frac{V R_{s}}{\lambda A} N = N( \frac{A}{{R_s}^2})^{-1} (\frac{V}{\lambda R_s}) $.

First, we can observe a clear performance dominance for the case where communication between AUVs is available over the case where it is not. {  The two branches (with and without communication) are both monotonically increasing from 0 to $100\%$, but cannot be simply collapsed onto each other by, for example, multiplying the horizontal axis of one branch by a constant. This is consistent with our argument at the beginning of Sec.~\ref{sec:scal-anal}, that different algorithms should in general produce different scaling functions. In this particular case, the difference between these two branches} highlights the categorical improvement that communication brings to this process. { To make this concrete, at $P_a = 20\%$ the no-communication branch requires roughly three times the value of $N_{\rm eff}$ needed by the communication branch, and at $P_a = 80\%$ that factor grows to roughly an order of magnitude. A single multiplicative rescaling of the horizontal axis therefore cannot bring the two branches together.} We also observe the break point for the case with communication capabilities around $N_{\text{eff}} = 1$ where the steep increase in performance breaks and is followed by a saturated behavior near the $100\%$ zone.
In contrast, the case where communication capabilities are not available has a more moderate increase { that continues after $N_{\text{eff}} = 1$. This curve will likely saturate eventually for very large numbers of drones, but at a much larger value of $N_{\text{eff}}$.}

Finally, we note that the battery life $B$ was chosen to be sufficiently large not to matter (i.e., $B$ was not included in the data scaling). If $B$ were decreased, we might expect the dimensionless ratio $BV(2R_s)N/A$ to play a role, which can be thought as the maximum coverage area per drone, $2BVR_s$, times the number of drones $N$ divided by the total area $A$ to be searched. When this dimensionless number is much greater than 1, as it is for nearly all our simulations, then battery life is not expected to play a role. As this ratio becomes close to 1 or less than 1, it may play a dominant role. To emphasize this effect, we set $B$ as an infinite number as presented in the results of this case in Fig.~\ref{fig:Pa_vs_Neff}, which shows a very similar performance to the {  simulations with a limited battery life, for which this dimensionless ratio is} still greater than {  $1$.}

\section{Case Study 3: Scattering Red Swarm}

\begin{figure} 
\centering
\includegraphics[width = 0.6\columnwidth]{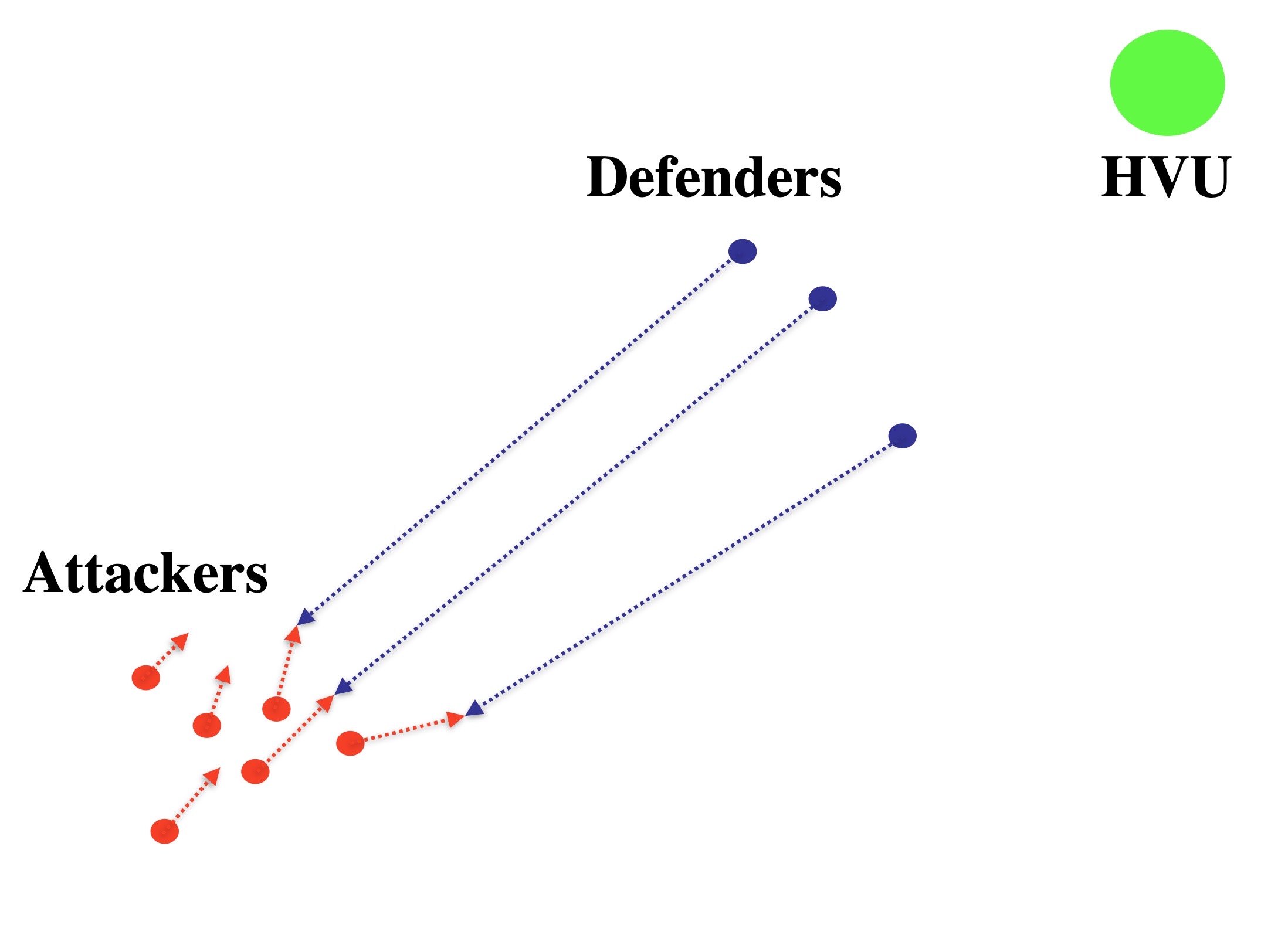}
\caption{A schematic of global targeting with intercept-based pursuit, where defenders fly toward an intercept point.}
\label{fig:intc} 
\end{figure}

Our final case study is motivated by the Service Academies Swarm Challenge (SASC), a recent live-fly swarm exercise involving the US military service academies~\cite{navy_drone,army_drone}. Each team was provided a swarm of up to 25 drones, including model weapon systems, with objectives including trying to destroy an opposing team's drones and control aerial {  territory~\cite{army_drone}.} Teams were provided with several drone targeting and control algorithms with varying features. One algorithmic feature was the targeting type: either local targeting, where each drone targeted the closest enemy, or global targeting, where targets were assigned by a global algorithm such that no two drones from the same swarm targeted the same enemy. Another feature involved the guidance law used to pursue the target (e.g., pursuit guidance, where $\mathbf{r}_{vl}$ is set to be the current position of the targeted enemy, versus an intercept-type guidance law, where $\mathbf{r}_{vl}$ is set at the point where the target enemy will be met). These features are illustrated in Fig.~\ref{fig:intc}. Unsurprisingly, global targeting and intercept-based guidance improved performance, and the number of drones in the swarm played a major role in mission success. However, this still leaves unanswered all the questions related to trade-offs raised in the introductory sections and the previous case study. The cost, including the sheer amount of man-hours, is prohibitively high to run live-fly exercises with wide variations in numbers of drones and drone platform specifications.

\subsection{Scaling Analysis using SASC algorithms}
\label{SASC_analysis}
Thus, in this section, we consider scaling and trade-off behavior for an algorithm similar to the most effective algorithm from the SASC, where both global targeting and intercept-based guidance are used. We assume a simple scenario, where an attacking red swarm of $N_a$ agents is attacking an HVU and senses a defending blue swarm of $N_d$ agents located a distance $D$ away. It then scatters, with each red agent moving with a constant random velocity $v_a^i$, uniformly distributed from $0<v_a^i<V_a$, with a constant random angle $\theta$, where $\theta=0$ denotes the direction toward the HVU and $\theta$ is uniformly distributed over the range $-\pi/4 < \theta < \pi/4$. The blue swarm has weapons with range {  $R_d$ and} is then tasked with visiting each red agent in the shortest time possible. 


This is accomplished by assigning $\mathbf{r}^{(j)}_{vl}$ for each defender {  $j$} in a way that corresponds to the attacker {  $i$} that it is pursuing (discussed further below), and then allowing the trajectory to evolve according to Eq.~\eqref{eqn:eq-of-mot-def}. The weapons model for the defenders is set to be
\begin{equation}
    { \phi_{ad}^{(il)}(\mathbf{r}_i-\mathbf{r}_l)} = \begin{cases}
        \infty &\text{, if }{ |\mathbf{r}_i-\mathbf{r}_l| < R_d}, \\
        0 &\text{, if }{ |\mathbf{r}_i-\mathbf{r}_l| \geq R_d},
    \end{cases}
    \label{eqn:DefenderWeapon}
\end{equation}
such that any attacker {  within a distance $R_d$ of a defender} is immediately killed. 

Targets for blue agents are chosen globally according to the following procedure. Attackers are assigned a priority level based on their distance from the point at which they scattered. The attacker with the highest threat level is assigned the closest defender to it at that instant in time. Next, the attacker with the second-highest threat level is assigned the closest defender to it at that time, except for the previously assigned defender. This process is repeated such that each attacker is targeted by the closest defender to its position from the pool of unassigned defenders. In the case where there are more defenders than attackers remaining, defenders are assigned to the closest remaining attacker. Thus, multiple defenders do target the same attacker, but only in the case where there are more defenders than attackers. We note that such an algorithm requires perfect communication among all defenders or perfect knowledge of the entire scenario, such that each defender can run the algorithm independently and know what all other defenders would be doing. This process is repeated at every time step, meaning target assignments can (and do) change before the current target is destroyed. 

Once the targeting assignments are made, the second stage of each algorithm involves setting $\mathbf{r}^{(j)}_{vl}$. We use something approximating proportional navigation, where the constant vector velocity of the targeted attacker and the maximum speed of the defender are used to calculate an intercept point, which we then assign to $\mathbf{r}^{(j)}_{vl}$. This assignment is also done at every time step, such that if a defender's velocity is not currently close to its maximum velocity, then $\mathbf{r}^{(j)}_{vl}$ will change with time. As the defenders velocity approaches its limiting maximum velocity, changes in $\mathbf{r}^{(j)}_{vl}$ with time become very small.

In this scenario, it is impossible for the attackers to ``win'' in the sense that the defenders will eventually catch them, so we always have $P_a = 0$ at the end of the simulation. Thus, as a performance metric, we measure the total kill time $t_k$ required for all attackers to be destroyed. That is, the probability of survival {  $Q_i$} of each attacker {  $i$, computed from Eq.~\eqref{eqn:prob-surv-att} using the attrition function of Eq.~\eqref{eqn:DefenderWeapon},} must be zero at $t_k$.  Each measurement of $t_k$ shown for the remainder of the paper is an average over ten simulations in order to mitigate the effect of outliers due to the random initial velocities of the attackers. The goal of our analysis is to obtain a functional form for the average $t_k$ in terms of all parameters that go into the simulation, including the numbers of drones $N_a$ and $N_d$ as well as the speed $V_d$ and weapons range {  $R_d$} of the defenders. The characteristic speed of attackers $V_a$ is held fixed, and we use the {  velocity ratio $\nu = V_d/V_a$} below.

\begin{figure}[h] 
\raggedright
(a) \\
\includegraphics[width = 0.99\columnwidth]{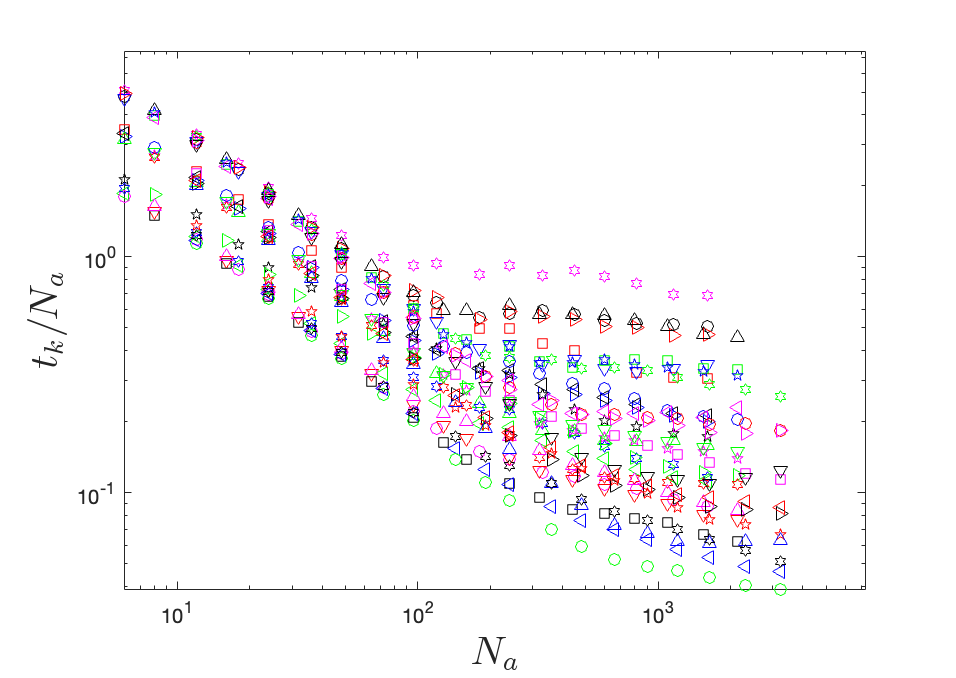} \\
(b) \\
\includegraphics[width = 0.99\columnwidth]{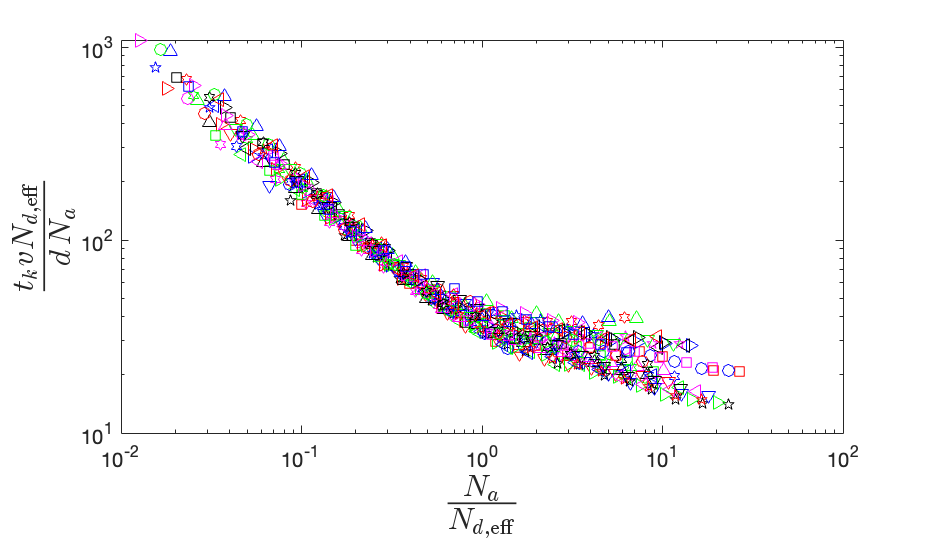}
\caption{(a) $t_k/N_a$ versus $N_a$ for simulations involving a large range of parameter values, with $V_a = 0.4$; $N_d = 24$, 32, and 48; {  $R_d = 1.2$} and 2.4; $\tau= 5$ and 10; and $V_d = 0.5$, 0.67, and 1. (b) The same data collapse when plotted as {  $t_k V_d N_{d,{\rm eff}}/(d N_a)$} versus $N_a / N_{d,{\rm eff}}$;{  the symbol $v$ in the axis label of the figure denotes the defender speed $V_d$.}}
\label{fig:alldata} 
\end{figure}

Data for $t_k$ is visualized by fixing all characteristics {  of} an engagement except {  $N_a$} and plotting the kill time per attacker, $t_k/N_a$, as a function of {  $N_a$.} Such curves are shown in Fig.~\ref{fig:alldata}(a) for many different values of $N_d$, {  $R_d$,} $\tau$ and $V_d$.


We note that all curves shown in Fig.~\ref{fig:alldata} have a similar shape: the kill time per attacker decreases with $N_a$ until a characteristic size of the attacking swarm, above which $t_k/N_a$ is approximately constant. This constant behavior corresponds to $t_k$ linearly proportional to the number of attackers. We define the value of $N_a$ that separates these two regimes as the effective size of the defending swarm, $N_{d,{\rm eff}}$. When $N_a<N_{d,{\rm eff}}$, the attacking swarm is ``effectively smaller'' than the defending swarm, and more attackers can be added without a proportional increase in $t_k$.  When $N_a>N_{d,{\rm eff}}$, the attacking swarm is ``effectively larger'' than the defending swarm, and adding more attackers yields a proportional increase in $t_k$.

We note that overall magnitude of these curves as well as the value $N_{d,{\rm eff}}$ depend on defender and attacker swarm parameters. A key question that mission planners would want to answer for this particular algorithm is how to stay in the first regime, where $N_a<N_{d,{\rm eff}}$ and the defending force has an advantage. Minimum resource allocation would correspond to analyzing an attacking swarm and deploying the right number of defenders such that $N_a \approx N_{d,{\rm eff}}$.


Using a similar procedure to the one used in the first case study, we use scaling analysis to search for a formula relating the performance metric $t_k$ to all system parameters. To form dimensionless groups, we consider $t_k$, $N_a$, $N_d$, $V_a$, $V_d$, the characteristic acceleration time $\tau$ for defenders, the defender weapon range {  $R_d$,} and a typical length scale $d$ for a total of eight parameters. We do not change any length scales in the system, so we assume $d\approx 1$. Since there are two dimensions (length and time) represented in these parameters, there should be six dimensionless groups, which can be written as $N_a$, $N_d$, {  $t_k V_a/d$, $\nu = V_d/V_a$, $\tau V_a/d$, and $R_d/d$.}



To obtain a collapse using scaling analysis, a similar procedure is used to the first case study. We first fit the logarithm of the data for $t_k/N_a$ versus $N_a$ using two linear functions, one for the downward sloping regime (where $N_a < N_{d,{\rm eff}}$) and one for the constant regime (where $N_a > N_{d,{\rm eff}}$). The intersection of these two lines is easily computed, which yields $N_{d,{\rm eff}}$ as a function of all dimensionless quantities. We vary each one systematically and find functions that approximately capture the dependence of each dimensionless quantity: we find that $N_{d,{\rm eff}}$ is linearly dependent on {  $R_d$,} depends on $N_d$ as a power law with exponent $3/2$, and is exponentially dependent on {  $\nu$} and $\tau$. Combining all, we obtain 
\begin{equation}
    N_{d,{\rm eff}} = {N_{d}}^{3/2}\left({ \frac{R_d}{d}} \right) { e^{\frac{4\nu}{5}}}e^{\frac{{\tau}{V_a}}{8 d}}.
    \label{eqn:N_d-eff}
\end{equation} 
Of particular interest is the nonlinear scaling in $N_d$, compared with a linear scaling in the first case study. The nonlinearity may be caused by the asymmetry of the competing swarms' goals (red swarm scatters, blue swarm chases).

We find that plotting $N_a / N_{d,{\rm eff}}$ on the horizontal axis approximately lines up all the crossover points. For the vertical axis, we seek to rescale $t_k/N_a$. First, instead of $t_k$, we use the dimensionless quantity {  $t_k V_d/d$.} We also postulate that $N_a$ should be replaced by $N_a / N_{d,{\rm eff}}$, suggesting that we should plot on the vertical axis the quantity {  $t_k V_d N_{d,{\rm eff}}/(N_a d)$.} Fig.~\ref{fig:alldata} shows this quantity plotted versus $N_a / N_{d,{\rm eff}}$, which collapses the data reasonably well. This suggests that
\begin{equation}
    t_k \approx \frac{N_a d}{N_{d,{\rm eff}} { V_d}} f\left(\frac{N_a}{N_{d,{\rm eff}}} \right)
\end{equation}
where the function $f$ can be roughly approximated as
\begin{equation}
    f(u) = \begin{cases}
         u^{-\alpha} &\text{, if } u<1, \\
        u^{-\beta} &\text{, if } u>1.
    \end{cases}
\end{equation}
The exponent $\alpha\approx 0.75$ captures nearly all the data when $N_a / N_{d,{\rm eff}}<1$. For $N_a / N_{d,{\rm eff}}>1$ the data are slightly more scattered, but $\beta\approx 0$ could likely be used with good fidelity. Further work might examine how $\beta$ depends on the dimensionless quantities in the system, but this is not addressed here since this study is primarily a demonstration of a method of analysis. Overall, the result of the analysis method we demonstrate here is a simple yet non-intuitive result that could only be obtained by a procedure like the one we show here.

{ To quantitatively evaluate the quality of this scaling collapse, we compute the Root Mean Squared Error (RMSE) of the compiled data points relative to a smooth, low-order polynomial curve fitted to the collapsed data in natural log-space. As a baseline for comparison, we establish a theoretical limit on the quality of the collapse by manually and independently shifting each individual curve to maximize their overlap (specifically, by forcing the inflection points between the small- and large-$N_a$ regimes to sit at the same point). This yields a minimum baseline RMSE of $0.123$ (representing a typical relative deviation for each data point of approximately $12\%$). The collapse we show in Fig.~\ref{fig:alldata}(b) yields an RMSE of $0.143$ (a typical relative deviation of $14\%$). Since the difference (RSME) in logarithmic space approximates percent error for small deviations ($\ln(y) - \ln(\hat{y}) \approx \frac{y-\hat{y}}{\hat{y}}$), this minor $2\%$ increase in relative scatter demonstrates the scaling collapse we show is nearly as good as possible. The minor remaining scatter in our model is primarily localized in the $u > 1$ regime, reflecting the curve-to-curve variations in the plateau exponent $\beta$ discussed above.}

\subsection{Scaling Analysis using Optimal Path Planning}
\label{Optimal_Scaling}
In the analysis introduced in the previous section every scenario studied required engaging every available defender. One may wonder, however, if this is really necessary, i.e., can the attackers be destroyed using a subset of the defenders available. In this section we demonstrate how the scaling analysis technique introduced in Section \ref{SASC_analysis} can be combined with optimal-path planning to determine the minimum number of defenders required to successfully destroy the attackers and to study the effectiveness of this approach. The details of the optimal-path-planning algorithm can be found in Appendix \ref{app:opt}.  

\begin{figure}
    \raggedright (a) \\ \centering
    \includegraphics[width=0.99\columnwidth]{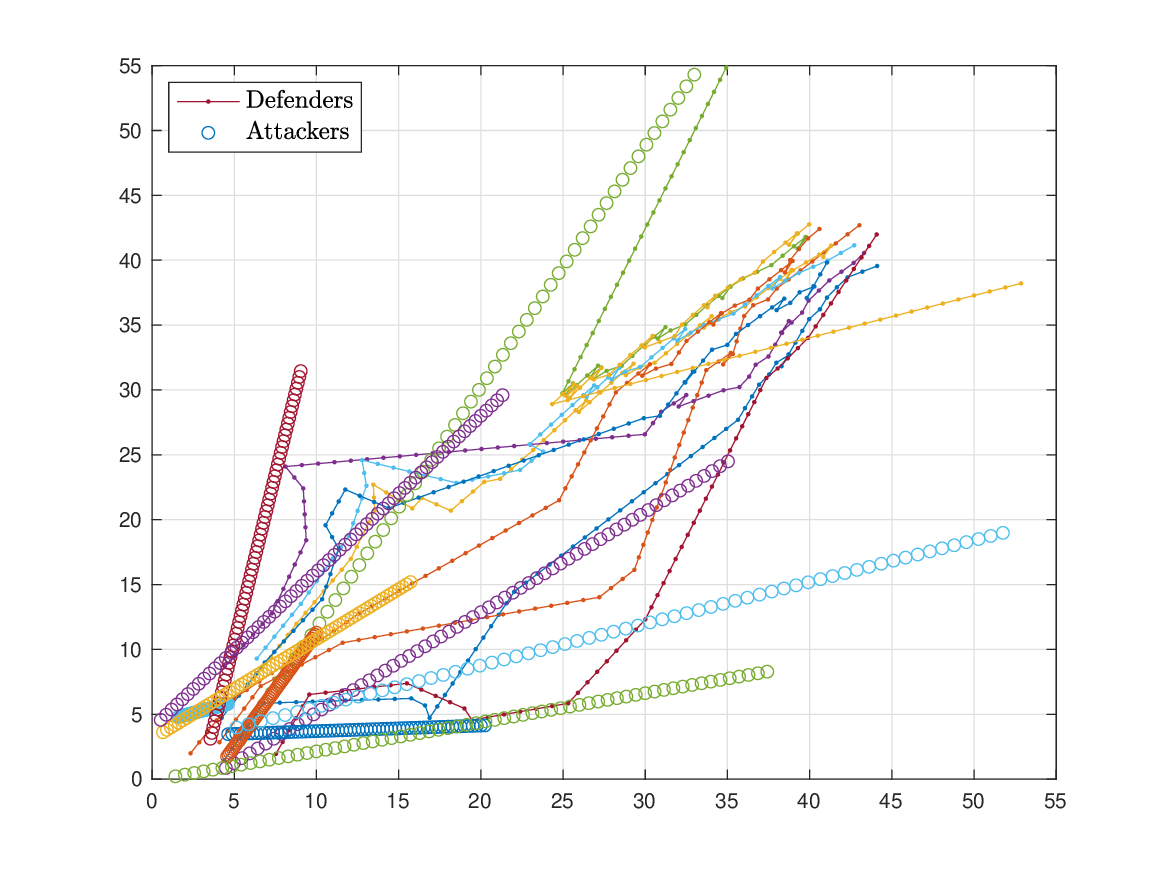}
    \\     \raggedright (b) \\ \centering
    \includegraphics[width=0.99\columnwidth]{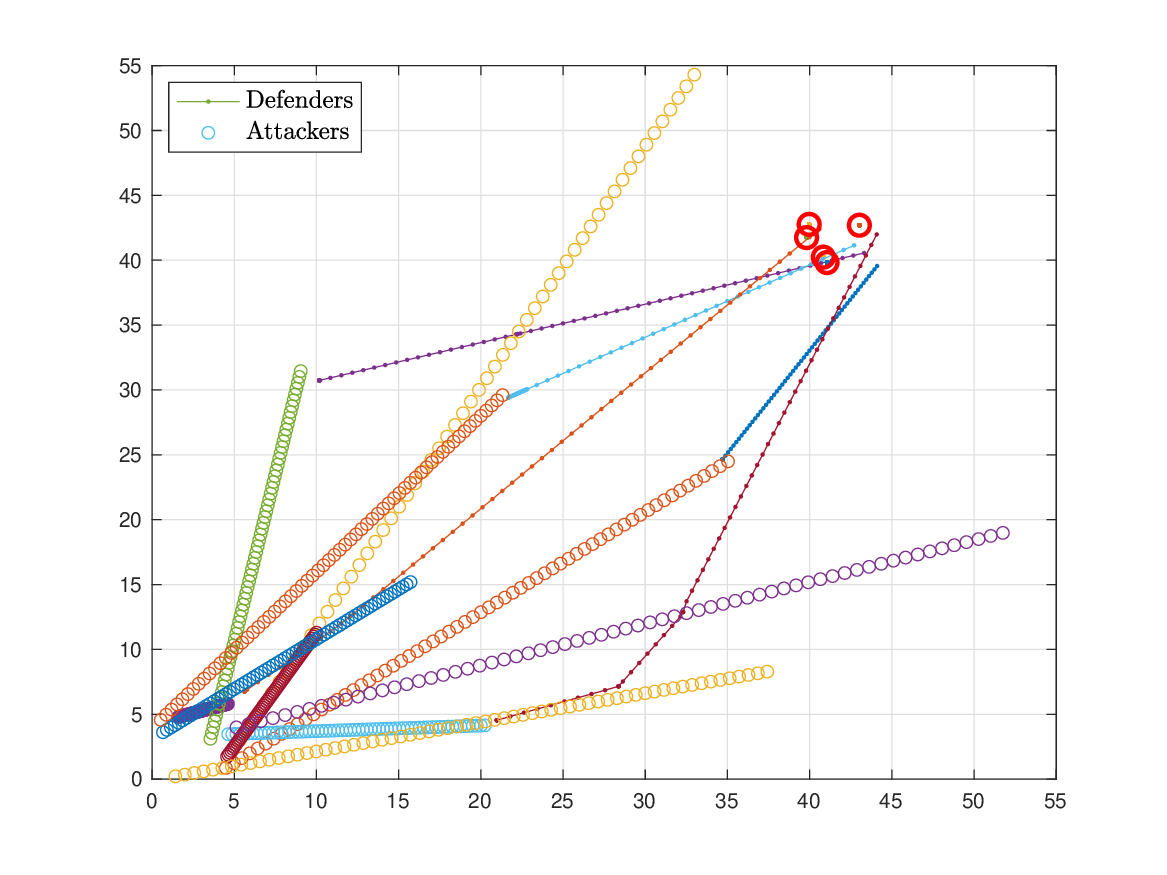}
    \caption{(a) The initial engagement, using global targeting with intercept-based pursuit, as shown in {  Fig.~\ref{fig:intc}.} Attackers' positions are traced out with open circles, and defenders are dots connected by lines. The color scheme is random, used to distinguish different agents. (b) The engagement after optimization, as described in the text. Red circles indicate defenders that were assigned to stay at the base by the optimization algorithm.}
    \label{fig:casadi_noOpts}
\end{figure}

The numerical solutions to the optimal path planning problem introduced in (\ref{eq:optimization-continuous} - \ref{eq:survive-constraint}) were initialized using defender trajectories developed in Section \ref{SASC_analysis}. The resulting solution used a subset of all the defenders available as shown in Fig.~\ref{fig:casadi_noOpts}. In fact, in this case, the off-the-shelf solution used all 10 defenders, whereas the optimal solution used only five.  We believe that the reduced number of defenders obtained is a direct consequence of the definition of the cost $J$ in (\ref{eq:optimization-continuous}) and the introduction of the restriction on the maximum probability of survival of each attacker. Indeed, since the cost $J$ is equal to the total distance traveled by all the defenders, using as few defenders as possible is a natural outcome of this choice. That this number is greater than zero is guaranteed by introducing a constraint on the maximum probability of survival of each attacker.
Moreover, the optimizer solved the assignment and sequencing problems implicitly. In other words, the targets assigned to each defender and the sequence by which they should be approached was never explicitly included in the optimal path planning formulation (\ref{eq:optimization-continuous} - \ref{eq:survive-constraint}).

\begin{figure}[h] 
\centering
\includegraphics[width = 0.99\columnwidth]{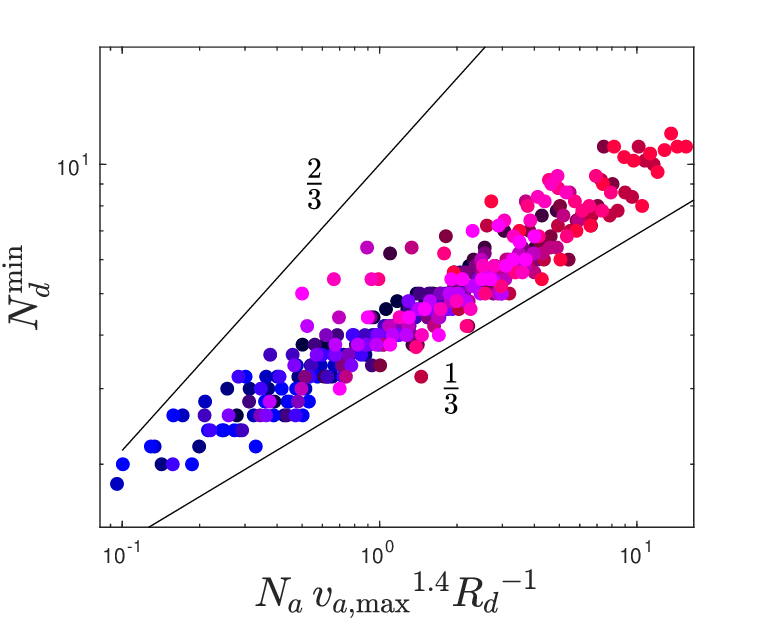} \\
\caption{Scaling of {  the minimum number of deployed defenders $N_d^{\rm min}$} with $N_a v_{a,\mathrm{max}}^{1.4} R_d^{-1}$. Marker color represents normalized attacker velocity $v_{a,\mathrm{max}}$ (blue to red) and defender range $R_d$ (darker shades indicate larger $R_d$).}
\label{fig:plot_velocity_sweep} 
\end{figure}

{  Finally, to compare with the scaling law in Eq.~\eqref{eqn:N_d-eff}, we repeat the calculation shown in Fig.~\ref{fig:casadi_noOpts} across the ranges $1\leq N_a \leq 40$, $0.25 \leq v_{a,{\rm max}} \leq 1$ (the maximum attacker speed, written $V_a$ above), and $0.1 \leq R_d \leq 10$, with the defender speed held at $1$. The results of these calculations are shown in Fig.~\ref{fig:plot_velocity_sweep}. Each point in Fig.~\ref{fig:plot_velocity_sweep} is the outcome of a separate optimization of the kind shown in Fig.~\ref{fig:casadi_noOpts}, so that Fig.~\ref{fig:plot_velocity_sweep} is a composite summary of the full sweep rather than a single engagement. }The outcome of each optimal engagement is the minimum number of defenders committed to neutralize all targets, denoted $N_d^{\rm min}$.
Using scaling analysis we have found that $N_d^{\rm min}$ increases with increasing $N_a$ and $v_{a,{\rm max}}$ and decreases with increasing $R_d$. These are consistent with physical intuition, i.e., more defenders are needed for more drones, for faster drones, and for a shorter range of the defenders' weapons, as shown in Fig.~\ref{fig:plot_velocity_sweep}.

{ As a preliminary observation, the number of deployed defenders $N_d^{\rm min}$ appears to scale approximately as $N_a^{1/3}$ over the range studied, compared with the $N_a^{2/3}$ dependence inferred from the completion-time crossover for the off-the-shelf algorithm. Specifically, the off-the-shelf case suggests that the scaling breakpoint occurs when $N_a/{N_{d,{\rm eff}}}\approx 1$, which by Eq.~\eqref{eqn:N_d-eff}, corresponds to $N_d^{\rm min} \approx N_a^{2/3}$.

These preliminary results are included to illustrate that the scaling framework can be used on optimized data. However, in addition, they also demonstrate that algorithms that include optimization may be able to dramatically improve performance. The change in the scaling form from power 2/3, inferred from the data shown in Fig.~\ref{fig:alldata}, to something closer to 1/3 represents a qualitative improvement in performance between the two algorithms. We contrast this with the performance curves in {  Fig.~\ref{fig:alldata}(a), which} all use different drone characteristics with the same {algorithm;} they are shifted by some constant but otherwise have the same scaling (i.e., the same slopes). We emphasize that optimization might significantly decrease computational efficiency and may not be possible in a real-time situation.}

\section{Conclusions}

This paper set out {  to} simplify the challenges related to the large design space of autonomous drone swarms by applying dimensional analysis and data‑scaling techniques. We executed over two million agent‑based simulations across three benchmark scenarios: a swarm‑on‑swarm battle, a cooperative area search subject to attrition, and a pursuit of scattering targets. In all cases, we used wide ranges of platform specifications, such as speed, sensing radius, weapon range, and attrition coefficients. By forming dimensionless groups from the governing variables, we showed that intuitive performance metrics could be collapsed onto a single, interpretable curve, expressed as an ``effective‑swarm‑size'' metric.
These curves allow us to identify the \textit{scaling break points} and obtain high-level findings that could be very useful to mission planners and design engineers. For example:
\begin{itemize}
    \item for the swarm-on-swarm battle, blue victory is achieved when the defender count satisfies $N_{d} \approx 2 N_{a,{\rm eff}}$, where $N_{a,{\rm eff}}$ can be written in terms of dimensionless ratios of blue and red agent characteristics.
    \item In search, {  full-coverage ($P_a \approx 100\%$)} is achieved efficiently when $\frac{VR_s}{\lambda A}(N-1)\geq 1$ {  and} limited inter‑agent communication is enabled; when no communication is present, {  full coverage does not} occur until at least $\frac{VR_s}{\lambda A}N > 10$.
    \item In pursuit, the ``effective-swarm-size'' metric predicts total time to chase down all agents. {  Preliminary optimal-path-planning results suggest that optimization may improve the scaling of the number of defenders required.}
\end{itemize}

In summary, we find that complex, seemingly mission-specific swarm behaviors can be analyzed with a single data-scaling framework. In all three cases we study, we find that the performance metric is governed by simple yet counterintuitive scaling laws that map the design space into regions of success and failure. While this method requires many simulations to be run over a wide range of parameters, these can be done offline and represent one-time cost for each scenario. The output is an analytical expression for performance, and evaluating these in real time is computationally trivial. 

{ We again emphasize that this work makes no claims to any universal scaling laws: all reported exponents and functional forms are specific to the models we study, which may or may not be good approximations of real-world drone engagements. For any specific set of ODEs that model an engagement, the scaling exponents and other details about the functional form (including confidence bounds) could be measured more precisely, in the event that is beneficial. We also note that since the exponent values are properties of the specific models studied, they may differ if the underlying model assumptions change, though the dimensional analysis procedure itself remains applicable regardless.} Similarly, the embedded optimal control results are included as a proof of concept; a systematic statistical characterization of exponent uncertainty and a formal convergence or robustness analysis are left for future work. Finally, future work might study both the ensemble-average, as we have done here, as well as the typical deviation from the ensemble average, which might be important when noise or randomness is dominant. Incorporating realistic communication impairments, such as packet loss, delays, or limited bandwidth, would introduce additional dimensionless groups into the framework and is also left for future work.

\appendix
\section{Optimal Path-Planning \label{app:opt}}
Let $S_d$ and $S_a$ denote the sets of defenders and attackers, with indices $j \in S_d$ and $i \in S_a$. Defender trajectories are denoted by $\mathbf{r}_j(t)$ over a free terminal time $T_{\mathrm{final}}$. The quantity $p^\mathrm{surv}_{ij}$ denotes the probability that attacker $i$ survives defender $j$, defined via the {  distance-dependent attrition rate $\rho_{ij}(t)$,} and $p^\mathrm{surv}_i$ denotes the total survival probability of attacker $i$ against all defenders.


The paths for the defenders $\mathbf{r}_j(t)$, $\forall j\in S_d$ are obtained by
solving the following multi-agent optimal path planning problem with
variable termination time $T_\mathrm{final}$, with given initial positions
$\mathbf{r}_j(0)$, $\forall j\in S_d$:
\begin{align}
  \label{eq:optimization-continuous}
  &\text{minimize}
  &&J:=\ \sum_{j\in S_d} \ell_j\\
  &\text{w.r.t}
  &&T_\mathrm{final}, \quad \mathbf{r}_j(t),\; \forall j\in S_d,t\in(0,T_\mathrm{final})\\
  &\text{subject to}
  &&\|\dot{\mathbf{r}}_j(t)\|\le v_{\max},\;\forall j\in S_d,t\in(0,T_\mathrm{final})\\
  &&&\|\ddot{\mathbf{r}}_j(t)\|\le a_{\max},\;\forall j\in S_d,t\in(0,T_\mathrm{final})\\
  \label{eq:survive-constraint}
  &&& p^\mathrm{surv}_i\le p^\mathrm{surv}_{\max},\;\forall i\in S_a,
\end{align}
where $\ell_j$, $j\in S_d$ denotes the path length for the $j$th defender
\begin{align*}
  \ell_j\eqdef \int_0^{T_\mathrm{final}} \|\dot{\mathbf{r}}_j(t)\| dt,
\end{align*}

which we use as a proxy for energy expenditure, since under bounded-speed
and bounded-acceleration motion, the mechanical work scales linearly with
traversed path length up to a constant factor; $p^\mathrm{surv}_i$, $i\in S_a$
denotes the probability that the $i$th attacker will survive all defenders, given by
\begin{align*}
  p^\mathrm{surv}_i&\eqdef \prod_{j\in S_d} p^\mathrm{surv}_{ij}, &
  p^\mathrm{surv}_{ij}&=e^{-\int_0^{T_\mathrm{final}} { \rho_{ij}(t)} dt},
\end{align*}
where $p^\mathrm{surv}_{ij}$ is the probability that attacker $i$ survives
defender $j$ and {  $\rho_{ij}(t)$} is the corresponding
distance-dependent {  attrition} rate
\begin{multline*}
   \Prob\big(\text{attacker $i$ { is destroyed by} defender $j$ on interval $(t,t+dt)$}\big)\\
  = { \rho_{ij}(t)} dt+ O(dt^2),
\end{multline*}
that we assume to be of the form
\begin{align*}
  { \rho_{ij}(t)}&\eqdef \lambda_{ij} \Phi\Big(\frac{F_{ij}-a_{ij} \|\mathbf{r}_i-\mathbf{r}_j\|^2}{\sigma_{ij}}\Big),
\end{align*}
where
$\Phi(z)\eqdef \int_{-\infty}^z \frac{1}{\sqrt{2\pi}}e^{-\xi^2/2}d\xi$ is the
cumulative function for the standard normal
distribution. {  Here $\lambda_{ij}$ sets the magnitude of the attrition rate,
$\sqrt{F_{ij}/a_{ij}}$ sets the distance at which that rate falls to
$\lambda_{ij}/2$, and $\sigma_{ij}$ controls how sharply it falls off with
distance. This form plays the same role as the attrition functions of
Eqs.~\eqref{eqn:prob-surv-att} and~\eqref{eqn:DefenderWeapon} in the main text,
but is written with a quadratic argument so that it is differentiable in
$\mathbf{r}_i-\mathbf{r}_j$, which the numerical solver requires.}

The numerical solution to \eqref{eq:optimization-continuous} relied on
a finite discretization of the (variable) horizon $(0,T_\mathrm{final})$
into $N_T$ intervals, corresponding collocation points
$t_0,t_1,\dots,t_{N_T}$ separated by $\delta t \eqdef
T_\mathrm{final}/N_T$. The velocities $\dot r$ and accelerations $\ddot
r$ were approximated by forward finite differences
\begin{align*}
  \dot{\mathbf{r}}_j(t)&\approx \frac{\mathbf{r}_j(t+\delta t)-\mathbf{r}_j(t)}{\delta t}, &
  \ddot{\mathbf{r}}_j(t)&\approx\frac{\dot{\mathbf{r}}_j(t+\delta t)-\dot{\mathbf{r}}_j(t)}{\delta t},
\end{align*}
and all the integrals by finite sums, as in
\begin{align}\label{eq:ell-j}
  \ell_j\approx { \sum_{k=0}^{N_T-1}}\|\dot{\mathbf{r}}_j(\delta t k)\| \delta t.
\end{align}
For numerical stability, the survivability constraints in
\eqref{eq:survive-constraint} were equivalently enforced in the
logarithm domain, as in
\begin{align*}
  \log p^\mathrm{surv}_i
  =-\sum_{j\in S_d} \int_0^{T_\mathrm{final}} { \rho_{ij}(t)} dt
  \le \log p^\mathrm{surv}_{\max},\quad\forall i\in S_a,
\end{align*}
with the integrals approximated by finite differences. In addition, to
avoid the lack of differentiability of the square-root needed to
compute the norms in \eqref{eq:ell-j}, we introduced slack variables
$\bar\ell_{j,k}$ subject to the (smooth) constraints
\begin{align}\label{eq:slack}
  &\bar\ell_{j,k} \ge 0, &
  &\bar\ell_{j,k}^2\ge { \delta t^2\,} \dot{\mathbf{r}}_j(\delta t k)' \dot{\mathbf{r}}_j(\delta t k),
\end{align}
and replaced the optimization cost in
\eqref{eq:optimization-continuous} by {  $\sum_{j\in S_d}\sum_{k=0}^{N_T-1} \bar\ell_{j,k}$.}
At the minimum, the slack variables $\bar\ell_{j,k}$ take the minimum
value allowed by the constraints in~\eqref{eq:slack}, which turns out
to be
\begin{align*}
  \bar\ell_{j,k}=\|\dot{\mathbf{r}}_j(\delta t k)\|{ \,\delta t},
\end{align*}
leading to the desired cost. The optimization solvers were generated
by Casadi \cite{AnderssonGillisHornRawlingsDiehl2018} using the IPOPT
interior-point solver \cite{BieglerZavaIPOPla2009}.



\begin{acknowledgements}

This research was funded by the Defense Advanced Research Projects Agency (DARPA) under grant number HR0011473967, by the Office of Naval Research, grant number N0001423WX01698, and by the Naval Research Program. The authors would like to thank Jeffrey Haferman for his help with high-performance computing at NPS.
This project was supported in part by an appointment to the NRC Research Associateship Program at the Naval Postgraduate School, administered by the Fellowships Office of the National Academies of Sciences, Engineering, and Medicine.
\end{acknowledgements}

\bibliography{thesis.bib}

\end{document}